\documentclass[fleqn,usenatbib]{mnras}

\usepackage{newtxtext,newtxmath}

\usepackage[T1]{fontenc}

\DeclareRobustCommand{\VAN}[3]{#2}
\let\VANthebibliography\thebibliography
\def\thebibliography{\DeclareRobustCommand{\VAN}[3]{##3}\VANthebibliography}

\usepackage{graphicx}	
\usepackage{amsmath}	
\usepackage{hyperref}
\usepackage{color}
\definecolor{patriarch}{rgb}{0.5, 0.0, 0.5}

\usepackage{natbib}
\usepackage{pdflscape}

\title[Nebular Attenuation Curves]{How Steep is your Dust? Direct Measurements of Nebular Attenuation Curves at Cosmic Noon with JWST}

\author[G. Maheson et al.]{Gabriel Maheson$^{1,2,3}$\thanks{E-mail: gabriel.maheson2@unibo.it},
Sirio Belli$^{3,4}$,
Sandro Tacchella$^{1,2}$,
Letizia Bugiani$^{3,4}$,
Joel Leja$^{5,6,7}$,
\newauthor Rebecca Davies$^{8,9}$,
Minjung Park$^{1,2}$ ,
Amir H. Khoram$^{3,4}$,
Matteo Sapori$^{3,4}$,
Caterina Liboni$^{3,4}$,
\newauthor Benjamin P. Collins$^{3,10}$
\\
$^{1}$Kavli Institute for Cosmology, University of Cambridge, Madingley Road, Cambridge CB3 0HA, UK\\
$^{2}$Cavendish Laboratory, University of Cambridge, 19 J. J. Thomson Ave., Cambridge CB3 0HE, UK \\
$^{3}$Dipartimento di Fisica e Astronomia, Università di Bologna, Via Gobetti 93/2, I-40129, Bologna, Italy\\
$^{4}$INAF, Astrophysics and Space Science Observatory Bologna, Via P. Gobetti 93/3, I-40129 Bologna, Italy\\
$^{5}$ Department of Astronomy \& Astrophysics, The Pennsylvania State University, University Park, PA 16802, USA\\
$^{6}$Institute for Computational \& Data Sciences, The Pennsylvania State University, University Park, PA 16802, USA\\
$^{7}$Institute for Gravitation and the Cosmos, The Pennsylvania State University, University Park, PA 16802, USA\\
$^{8}$Centre for Astrophysics and Supercomputing, Swinburne University of Technology, Hawthorn, Victoria, Australia\\
$^{9}$5ARC Centre of Excellence for All Sky Astrophysics in 3 Dimensions (ASTRO 3D), Australia\\
$^{10}$Departamento de Astronomía, Universidad de Chile, Camino El Observatorio 1515, Las Condes, Santiago, Chile\\
}

\date{Accepted XXX. Received YYY; in original form ZZZ}

\pubyear{2026}

\begin{document}
\label{firstpage}
\pagerange{\pageref{firstpage}--\pageref{lastpage}}
\maketitle

\begin{abstract}
Tracing the shape of the nebular attenuation curve, which is crucial for accurately measuring intrinsic emission-line fluxes, has proven challenging both in the local universe and at high redshift. Here we present nebular dust attenuation curves measured directly for 24 star-forming galaxies ($9.3<\log(\rm M{\star}/M_\odot)<11.1$) at cosmic noon ($1.8<z<3.3$) from the Blue Jay survey. Using deep JWST/NIRSpec spectra, we measure up to eight hydrogen recombination lines from H$\delta$ to Pa$\beta$ per galaxy and find little diversity in the curve shape, with most galaxies agreeing with the \citet{calzetti_dust_2000} attenuation curve. For over $90\%$ of the sample, the nebular reddening from the Balmer lines is consistent with that from the Paschen lines. Using \texttt{Prospector} spectral energy distribution (SED) fits, we obtain the stellar reddening and find that the difference between stellar and nebular reddening increases with stellar mass, and that each reddening measurement is independent of inclination, which disagrees with the two-component dust model. We then measure the impact of using the derived nebular attenuation curve or a literature curve when determining metallicity using the strong-line method, and find that for some diagnostics (e.g., S32, S3, S23, and S3O3), the difference can be as large as $5\sigma$. Caution is needed when using these diagnostics if the nebular attenuation curve cannot be directly measured, since the resulting uncertainty can exceed the measurement uncertainty on the emission-line fluxes even with high signal-to-noise Balmer decrement measurements. These results show that the shape of the nebular attenuation curve is poorly understood and that constraining it is important for reducing systematic uncertainty in galaxy measurements.
\end{abstract}




\section{Introduction}

The luminosities and ratios of rest-frame optical emission lines are used to measure many key properties of galaxies, including the star formation rate \citep[SFR; e.g.,][]{kennicuttFutureStarFormation1994} and gas-phase metallicity \citep[see review by][]{maiolinoReMetallicaCosmic2019}. The photons producing these emission lines are scattered and absorbed by interstellar dust, a process known as dust attenuation \citep[see review by][]{salim_dust_2020}. Accurate modelling of this attenuation is therefore crucial for measuring galaxy properties, such as star-formation rate and gas-phase metallicity via emission-line measurements, and stellar mass and star formation histories (SFH) via spectral energy distribution (SED) fitting. 

The variation in dust attenuation with wavelength is known as the dust attenuation curve, and measurements of the dust curve have been made for decades. When measuring the dust attenuation of emission lines in local galaxies, \cite{fanelli_spectral_1988} and \cite{calzetti_dust_1994} found that emission lines often have higher dust attenuation than the overall stellar continuum. These observations led to the two-component dust model from \cite{charlot_simple_2000}, wherein all the stars are attenuated by a component of dust in the ISM, and the young stars, typically with ages $\lesssim10 \rm ~Myr$ \citep{blitz_origin_1980}, and the nebular emission lines produced around those stars, are additionally attenuated by the dust in the nascent birth cloud. After $\sim \rm 10 ~Myr$, these birth clouds disperse. Observations at higher redshifts have also identified elevated nebular reddening compared to that of the stellar continuum \citep{wuytsCANDELS3DHSTSynergyResolved2013, price_direct_2014, reddy_mosdef_2015, shivaei_mosdef_2020, lorenz_updated_2023}. 

Different geometries and optical depths toward young and old stellar regions also motivate different attenuation curves, since variation in attenuation curves tends to be dominated by geometry \citep[e.g.,][]{narayanan_theory_2018, salim_dust_2020}, with subdominant contributions from extinction curves. It is typical to assume that the dust in the birth clouds can be modelled as a foreground screen of dust with negligible scattering into the line of sight. Since this aligns with how the dust in the Milky Way is modelled \citep{cardelli_relationship_1989}, the shape of the Milky Way extinction curve is commonly assumed for the nebular attenuation curve. However, for the stellar continuum, dust scattering into the line of sight by ISM dust is a significant effect. Attenuation curves that model this scattering into the line of sight, such as the \cite{calzetti_dust_2000} law, are commonly assumed for the stellar continuum attenuation curve. 

Various methods have been used to measure the shape of the stellar continuum attenuation curve, such as the empirical template method by \cite{calzetti_dust_1994}, which stacks the spectra of local galaxies in bins of dust attenuation to infer the shape of the dust attenuation curve. More recently, \cite{reddy_mosdef_2015} and \cite{shivaei_mosdef_2020} applied this method to cosmic noon galaxies, and the attenuation curves measured in these three works vary in shape. Many studies have used SED fitting tools to infer the shape of the attenuation curve for local galaxies \citep{burgarellaStarFormationDust2005, leja_deriving_2017, salim_dust_2018} and for galaxies at cosmic noon \citep{mahesonBigDustyGalaxies2025} and at even higher redshifts \citep{markovUnveilingTrendsDust2025a, shivaeiDiversityEvolutionDust2025}. These works find significant variation in the shape of the stellar continuum attenuation curve, and that the slope and UV-bump strength depend on parameters such as dust content, stellar mass, SFR, and redshift.

However, the shape of the nebular attenuation curve has not been studied as widely, particularly at high redshift. Recent works using the Balmer emission lines to measure the nebular attenuation curve directly have shed light on its shape \citep{reddy_mosdef_2020, rezaeeVariationNebularDust2021, sanders_aurora_2024, reddyAURORASurveyMultiple2026a}. The Balmer lines are widely used to calculate the optical depth towards nebular regions \citep[e.g.,][]{garn_predicting_2010, shivaeiMOSDEFSurveyDissecting2015, maheson_unravelling_2024, sandles_jades_2024, mahesonBigDustyGalaxies2025}, since the intrinsic ratio of Balmer lines is well predicted theoretically, and varies little with the physical conditions they originate in \citep[e.g.,][]{tacchella_confinement_2016, chakraborty_x-ray_2021}. \cite{reddy_mosdef_2020} introduced a method to measure the nebular attenuation curve directly using several Balmer emission lines. They stacked galaxy spectra to measure the Balmer lines from H$\alpha$ to H$\epsilon$ at $z\sim1-2$. The low signal-to-noise of their pre-JWST spectra required stacking to boost the signal of the higher-order lines, allowing them to measure an average attenuation curve across this redshift range. This nebular attenuation curve was similar to the Milky Way extinction curve in the rest-optical. A similar result was reported for a stack of local ($z\sim0$) galaxy spectra using the same methodology \citep{rezaeeVariationNebularDust2021}, which found no dependence of the nebular attenuation curve on stellar mass, metallicity, or specific star-formation rate (sSFR).

Since the advent of JWST and the high sensitivity and multiplexing of NIRSpec, it has been possible to apply this methodology to individual galaxies, and more recently to samples of individual galaxies. So far, this has only been achieved using galaxies from the AURORA survey \citep{shapleyAURORASurveyNew2025}, first for an individual galaxy in \citet{sanders_aurora_2024}, and then in a follow-up paper \citep{reddyAURORASurveyMultiple2026a} for 24 galaxies. Both works find many galaxies with nebular attenuation curves far steeper than the Milky Way, SMC, or \citet{calzetti_dust_2000} attenuation curves, which are commonly used to dust-correct emission lines, as well as the curve reported by \citet{reddy_mosdef_2020} at similar redshifts. 

The curves reported by \citet{reddyAURORASurveyMultiple2026a} also show strong diversity, with the slope varying significantly from galaxy to galaxy, yet it is still unclear what causes this diversity, or whether we can constrain or predict it. In \citet{reddyAURORASurveyMultiple2026a}, they associate this diversity with variable dust covering fractions, stating that complex star-dust geometries could lead to a non-unity dust covering fraction, allowing some light from young stars (in the UV) to escape unattenuated and flattening the nebular attenuation curve. They argue that variation in this dust-covering fraction from galaxy to galaxy drives the diversity we see.
 
Beyond informing us about possible dust-grain properties or star-dust geometry in these galaxies, the shape of the attenuation curve has important implications for how we dust-correct emission lines used to measure the SFR (from H$\alpha$), gas-phase metallicity, and many other properties. Comparing the measured gas-phase metallicity for their galaxy in \cite{sanders_aurora_2024} using either the Milky Way (MW) extinction curve or their derived nebular attenuation curve, there is a discrepancy of $\sim0.2$~dex. Following a similar methodology, \citet{reddyAURORASurveyMultiple2026} show that the discrepancy in H$\alpha$ flux can be on the order of $0.2$ dex for the sample of attenuation curves measured in \citet{reddyAURORASurveyMultiple2026a}. With deep JWST/NIRSpec observations, measurement uncertainties on emission lines can now be smaller than the uncertainty introduced by the shape of the attenuation curve used to dust-correct the lines. Constraining the shape of the nebular attenuation curve at high redshifts is therefore timely and important, since directly measuring the nebular attenuation curve for each galaxy is not an option for most studies. 

In this work, we directly measure the nebular attenuation curves for a sample of 24 typical star-forming galaxies at $1.8<z<3.3$ observed with JWST/NIRSpec from the Blue Jay survey \citep{belliBlueJaySurvey2025}. Using this sample, we can assess the shape of the nebular attenuation curve galaxy by galaxy and compare our results with previous studies to identify what may drive any variation. We then leverage \texttt{Prospector} SED fits of these galaxies to measure their stellar population properties, compare the nebular and stellar reddening, and explore how the choice of attenuation curve affects the measured gas-phase metallicity for several different strong-line diagnostics. 

This paper is split into eight sections, with an overview of the data sample in Section~\ref{s:data_sample}, the method of measuring and fitting the nebular attenuation curve is outlined in  Section~\ref{s:calc_atten}, and the calculated nebular attenuation curves and the best fits to their functional forms are presented in Section~\ref{s:fitting_atten}. The resultant nebular attenuation curves are compared to other attenuation curves in Section~\ref{s:comp_studies}, a comparison between the nebular and stellar reddening is made in Section~\ref{s:comp_stellar}, and the impact of the choice of nebular attenuation curve on the measurement of the gas-phase metallicity is explored in Section~\ref{s:metallicity}. We conclude our results in Section~\ref{s:conclusion}. Throughout this paper, a cosmology with $H_0=70$ km s$^{-1}$ Mpc$^{-1}$, $\Omega_{\Lambda}=0.7$, and $\Omega_{m}=0.3$ is adopted.

\section{Data Sample}\label{s:data_sample}

The galaxies used in this work are a subsample of the galaxies from the Blue Jay survey \citep{belliBlueJaySurvey2025}. In Section~\ref{s:blue_jay_sample}, we provide an overview of the observations and discuss the stellar population modelling in Section~\ref{s:sed_fit} and the emission line fitting in Section~\ref{s:emission_fit} applied to the full sample. In Section~\ref{s:sample_selection}, we discuss how the sample of galaxies used in this work is selected.

\subsection{Blue Jay survey}\label{s:blue_jay_sample}

The Blue Jay survey consists of 153 galaxies from the JWST Cycle 1 program (GO 1810; PI Belli) spread across two pointings in the COSMOS field, forming a representative sample of galaxies at these redshifts. These galaxies were observed with the NIRSpec micro-shutter assembly (MSA; \citealt{ferruit_near-infrared_2022, rawle_-flight_2022}) to obtain $\rm R\simeq1000$ spectra with rest-frame coverage of $3700$~\AA{}-$1.2~\mu$m. Four of these galaxies were filler targets at $z\sim6$, leading to 149 galaxies with stellar masses in the range $8.6<\log(\rm M_{\star}/\rm M_{\odot})<12.0$ at cosmic noon ($1.6<z<3.7$). 

The NIRSpec observations for the Blue Jay sample utilised three medium-resolution gratings (G140M, G235M and G395M) with exposure times of 13h, 3.2h and 1.6h, respectively. Slitlets were constructed with at least two shutters for each galaxy, and an A-B nodding pattern was employed. A modified version of the JWST Science Calibration Pipeline v1.17.1 and the Calibration Reference Data System version 1321 are used to reduce the observed spectra for analysis. We derive a master sky spectrum from empty shutters for background subtraction. 

We use public NIRCam imaging in the F090W, F115W, F150W, F200W, F277W, F356W, F410M and F444W filters from the JWST Cycle 1 program PRIMER (GO 1837; PI Dunlop). We also calculated photometry using our NIRSpec slit aperture from publicly available HST/ACS+WFC3 \citep{skelton_3d-hst_2014, momcheva_3d-hst_2016} imaging in the F125W, F140W, F160W, F606W, and F814W filters, extending our photometric wavelength range into the UV. These photometric data are available at https://zenodo.org/records/17333505.

The full description of the Blue Jay survey, sample selection and data reduction steps is provided in the survey paper \citep{belliBlueJaySurvey2025}. The Blue Jay survey has given insight into the nature of massive cosmic noon galaxies, with \citet{belli_star_2024}, \citet{bugianiActiveGalacticNucleus2025}, \citet{davies_jwst_2024} and \citet{liboniProbingNeutralOutflows2026} studying the properties of their outflows and AGN, \citet{park_widespread_2024} studying their quenching properties, and \citet{mahesonBigDustyGalaxies2025} studying their dust attenuation and morphological properties. 

\subsection{Stellar Population Modelling}\label{s:sed_fit}

We use the Bayesian stellar population inference code \texttt{Prospector} \citep{johnson_stellar_2021} to fit the spectra and photometry (from the slit aperture) simultaneously to infer the stellar population properties of the galaxies in the Blue Jay survey. For details on how this matched-aperture SED fitting was performed, we refer the reader to Bugiani et. al. (in prep). Here, we summarise the methodology as follows: we adopt the synthetic stellar population library FSPS \citep{conroy_propagation_2009, conroy_propagation_2010}, the MIST isochrones \citep{choi_mesa_2016}, the C3K spectral library \citep{cargile_minesweeper_2020}, and assume the Chabrier initial mass function \citep{chabrier_galactic_2003}. The star formation history (SFH) was modelled using the non-parametric method from \cite{leja_how_2019, leja_older_2019}, with SFH bins logarithmically spaced in age. We then obtain the best-fit stellar population properties, such as the stellar mass (M$_{\star}$) and the SFH. In this study, we use the SFR from the most recent time bin, which covers $0-30$ Myr. 

The dust content of these galaxies is accounted for in \texttt{Prospector} through dust absorption and re-emission, which is assumed to be in energy balance. The dust attenuation is modelled following the two-component dust model, with a primary dust component which applies to all stars and follows the \cite{kriek_dust_2013} attenuation curve, as well as a second dust component which only attenuates young stars ($<10$ Myr) and follows a power-law attenuation curve with fixed slope. We extract the stellar continuum attenuation curve for each galaxy by dividing the model spectrum with dust "turned on" by the model spectrum with dust "turned off".

How this attenuation curve behaves with different galaxy properties for the full Blue Jay sample is explored in depth in \cite{mahesonBigDustyGalaxies2025}, where the stellar mass and SFR are shown to correlate strongly with the amount of dust attenuation ($A_{\rm V}$) and less so with the slope or UV-bump strength; however, the UV-bump strength and the attenuation curve slope are tightly correlated with $A_{\rm V}$.

\subsection{Emission Line Fitting}\label{s:emission_fit}

We calculated the emission line fluxes for the full parent sample, and the details and full catalogue are presented in Bugiani et. al. (in prep.); we summarise the process here. We subtract the best-fit stellar continuum derived through the \texttt{Prospector} fit, as discussed in the previous section, from the calibrated spectrum. We then fit each emission line with a Gaussian profile, with the velocity dispersion and redshift identical for all lines. When determining the velocity dispersion, we account for the contribution of NIRSpec's nominal wavelength-dependent spectral resolution to the overall line broadening. We use the Python library \texttt{EMCEE}, an Affine Invariant Markov Chain Monte Carlo (MCMC) ensemble sampler \citep{foreman-mackey_emcee_2013}, to fit the emission lines. When fitting, the only free parameters are the global redshift, velocity dispersion and the amplitude of each emission line. 

\begin{figure*}
\centerline{\includegraphics[width=.9\textwidth]{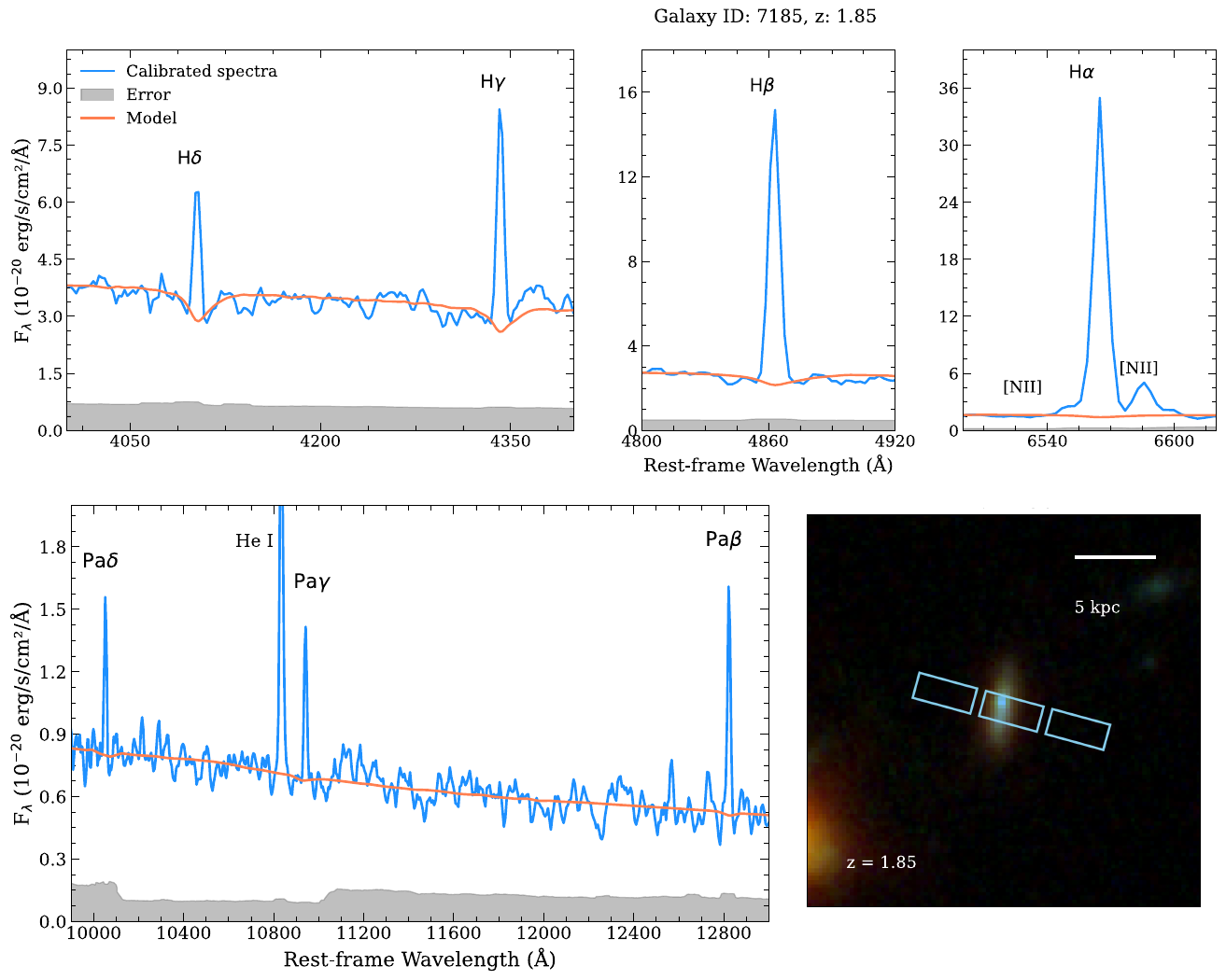}}
\medskip
\caption{Example JWST/NIRSpec spectrum of galaxy with ID 7185 used in this work shown in blue and the stellar continuum best fit shown in orange. The noise spectrum is shown as the grey curve at the bottom of each plot. This galaxy has seven Balmer and Paschen lines with a signal-to-noise ratio of at least 5 and was therefore selected for this work. In the bottom right, we show the RGB cutout of this galaxy in the F115W, F200W and F444W filters, with the slit footprint overlaid, along with the 5kpc scale bar for reference.} \label{f:example_spectra}
\end{figure*}

Various emission lines were measured for each galaxy in the sample for use in this work, namely the hydrogen recombination lines and strong lines. In Section~\ref{s:sample_selection} we discuss which hydrogen emission lines were used to measure the nebular attenuation curve for each galaxy, and in Section~\ref{s:metallicity} we discuss which strong lines were used to determine the metallicity for each galaxy. To showcase the quality of the spectra used in this work, we show the 1D spectrum of the galaxy with ID 7185 in Figure~\ref{f:example_spectra}. 

\subsection{Sample Selection}\label{s:sample_selection}

\begin{figure*}
\centerline{\includegraphics[width=.9\textwidth]{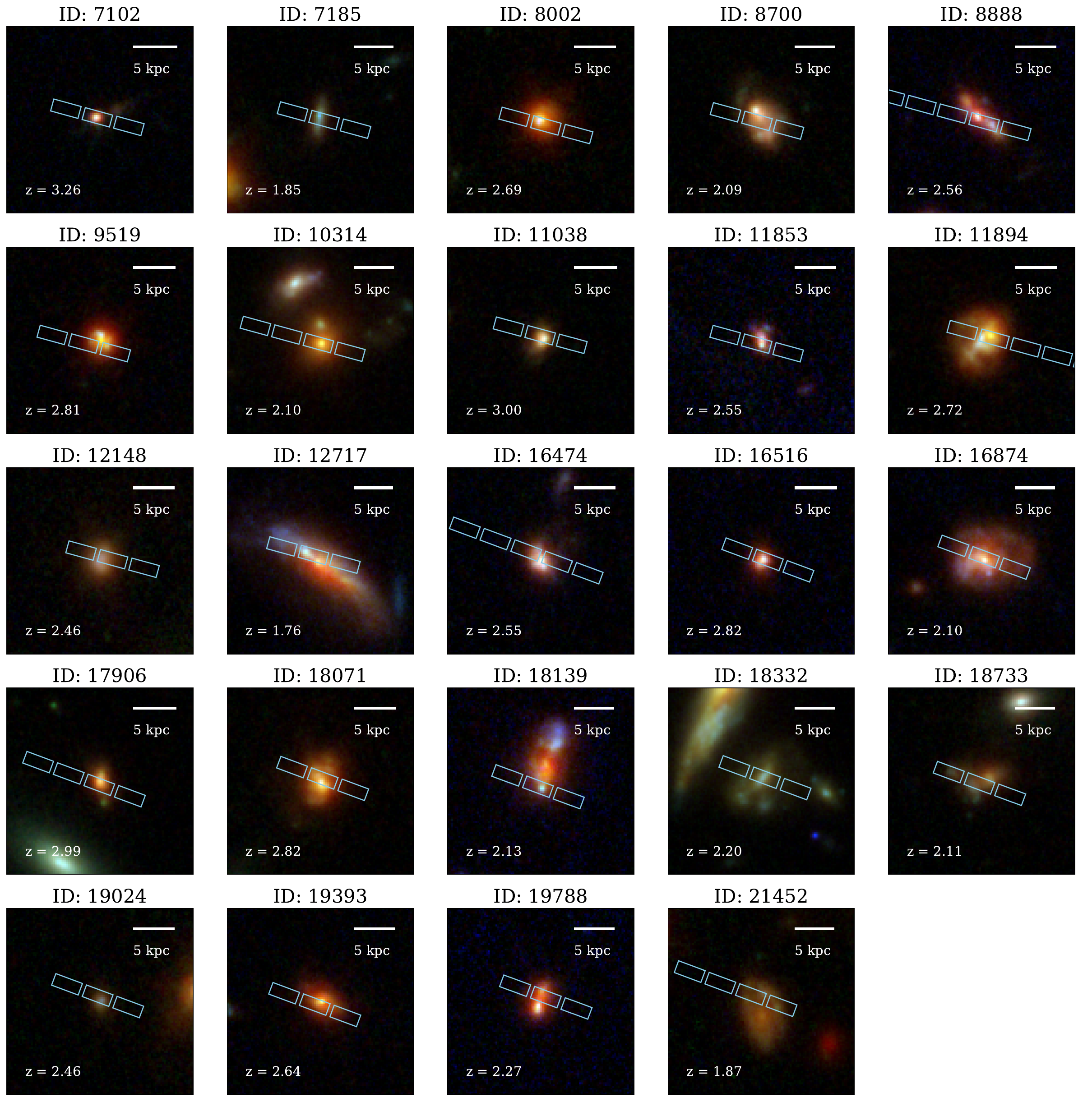}}
\medskip
\caption{RGB stamps using NIRCam F115W, F200W, F444W imaging of the 24 galaxies used in this work with the slits overplotted. Each slit is shown as a series of blue rectangles representing the NIRSpec micro-shutters. The redshift and a $5$kpc scale bar are also overplotted for reference.} \label{f:stamps_grid}
\end{figure*}

To measure the nebular attenuation curve, we require several high signal-to-noise detections of hydrogen recombination lines. Our rest-frame wavelength coverage allows for the entire Balmer and Paschen series (except Pa$\alpha$) to be covered. From the full Blue Jay sample, we select a subsample by setting a signal-to-noise cut of 5$\sigma$ and requiring at least 5 hydrogen recombination lines; similar selection criteria were also employed by \citet{reddyAURORASurveyMultiple2026a}. We additionally enforced the line ratio between H$\alpha$ and all other lines to obey Case-B recombination, leading to 7 galaxies being discarded. These selection criteria left us with a sample of 24 galaxies, and the emission line fluxes are shown in Table~\ref{t:EL_fluxes} for each galaxy. We did not consider the H$\epsilon~\lambda3971$, H$8~\lambda3890$ and Pa$8~\lambda 9546$ emission lines, since these are blended with the [Ne III]$~\lambda3970$, He I$~\lambda3890$ and [SIII]$~\lambda9533$ emission lines, respectively, and with our medium-resolution NIRSpec spectra ($\rm R\sim1000$), we can not confidently deblend these emission lines. We also set a $5\%$ error floor on all emission-line fluxes to ensure we do not underestimate the flux errors.

\begin{table*}
  \centering
    \begin{tabular}{cccccccccc}\hline\hline
        ID &  & H$\delta$ & H$\gamma$ & H$\beta$ & H$\alpha$ & Pa9 & Pa$\delta$ & Pa$\gamma$ & Pa$\beta$  \\\hline
        & $\lambda_{\rm rest}$ [\AA] & 4102 & 4340 & 4861 & 6563 & 9229 & 10050 & 10939 & 12819 \\ \hline
        7102 & \ & $96 \pm 4$& $169 \pm 4$& $432 \pm 11$& $1440 \pm 8$& $-$& $32 \pm 3$& $70 \pm 4$& $-$ \\
7185 & \ & $77 \pm 9$& $129 \pm 9$& $294 \pm 8$& $948 \pm 6$& $-$& $28 \pm 4$& $33 \pm 2$& $76 \pm 6$ \\
8002 & \ & $-$& $99 \pm 9$& $253 \pm 7$& $1622 \pm 12$& $-$& $-$& $131 \pm 7$& $291 \pm 7$ \\
8700 & \ & $147 \pm 17$& $301 \pm 19$& $681 \pm 18$& $2242 \pm 28$& $-$& $-$& $93 \pm 14$& $151 \pm 11$ \\
8888 & \ & $83 \pm 10$& $177 \pm 10$& $393 \pm 8$& $1501 \pm 13$& $-$& $-$& $68 \pm 8$& $125 \pm 8$ \\
9519 & \ & $-$& $101 \pm 12$& $282 \pm 10$& $1261 \pm 16$& $-$& $-$& $60 \pm 7$& $128 \pm 8$ \\
10314 & \ & $-$& $-$& $457 \pm 13$& $2949 \pm 20$& $76 \pm 9$& $113 \pm 5$& $325 \pm 11$& $624 \pm 9$ \\
11038 & \ & $67 \pm 9$& $131 \pm 7$& $371 \pm 14$& $1190 \pm 13$& $-$& $-$& $-$& $83 \pm 8$ \\
11853 & \ & $61 \pm 9$& $132 \pm 10$& $271 \pm 7$& $901 \pm 11$& $-$& $-$& $-$& $99 \pm 8$ \\
11894 & \ & $86 \pm 11$& $181 \pm 12$& $424 \pm 11$& $1601 \pm 32$& $-$& $-$& $-$& $159 \pm 19$ \\
12148 & \ & $281 \pm 24$& $570 \pm 22$& $1274 \pm 21$& $4502 \pm 29$& $-$& $127 \pm 17$& $243 \pm 18$& $386 \pm 16$ \\
12717 & \ & $-$& $-$& $328 \pm 25$& $2017 \pm 20$& $-$& $227 \pm 22$& $120 \pm 12$& $265 \pm 18$ \\
16474 & \ & $100 \pm 13$& $203 \pm 14$& $414 \pm 10$& $1804 \pm 17$& $-$& $-$& $89 \pm 10$& $-$ \\
16516 & \ & $-$& $97 \pm 8$& $217 \pm 5$& $1195 \pm 11$& $-$& $-$& $64 \pm 6$& $120 \pm 5$ \\
16874 & \ & $-$& $136 \pm 20$& $272 \pm 15$& $1337 \pm 22$& $-$& $-$& $77 \pm 11$& $131 \pm 9$ \\
17906 & \ & $56 \pm 7$& $103 \pm 6$& $323 \pm 11$& $1192 \pm 10$& $-$& $-$& $54 \pm 5$& $125 \pm 6$ \\
18071 & \ & $85 \pm 13$& $174 \pm 14$& $508 \pm 10$& $3006 \pm 18$& $-$& $90 \pm 8$& $209 \pm 9$& $414 \pm 9$ \\
18139 & \ & $-$& $89 \pm 7$& $245 \pm 7$& $1450 \pm 145$& $-$& $68 \pm 2$& $143 \pm 7$& $261 \pm 6$ \\
18332 & \ & $79 \pm 13$& $191 \pm 14$& $475 \pm 11$& $1988 \pm 17$& $-$& $-$& $65 \pm 7$& $134 \pm 6$ \\
18733 & \ & $204 \pm 17$& $415 \pm 16$& $911 \pm 14$& $3609 \pm 23$& $-$& $71 \pm 5$& $142 \pm 10$& $284 \pm 9$ \\
19024 & \ & $88 \pm 10$& $176 \pm 10$& $361 \pm 9$& $1120 \pm 12$& $-$& $-$& $39 \pm 5$& $77 \pm 5$ \\
19393 & \ & $-$& $73 \pm 9$& $208 \pm 8$& $1108 \pm 13$& $-$& $41 \pm 7$& $89 \pm 7$& $-$ \\
19788 & \ & $-$& $41 \pm 6$& $142 \pm 7$& $1209 \pm 12$& $33 \pm 4$& $-$& $163 \pm 7$& $337 \pm 8$ \\
21452 & \ & $-$& $137 \pm 17$& $373 \pm 14$& $1770 \pm 26$& $-$& $-$& $117 \pm 7$& $224 \pm 14$ \\
    
    \hline
    \end{tabular}\\[5pt]
    \medskip
    \caption{Table showing the hydrogen recombination line fluxes (in units of $\rm 10^{-18} ~ erg ~ cm^{-2} ~ s^{-1}$) selected for the galaxies used in this study. Many of the Balmer and Paschen lines that are not recorded in this table were measured for each galaxy but were not selected due to too low signal-to-noise. These emission lines are available in Bugiani et. al. (in prep).}
    \label{t:EL_fluxes}
\end{table*}

To show the diversity in galaxy shapes in this subsample, we show cutouts of these 24 galaxies in Figure~\ref{f:stamps_grid}; some appear compact with a clear disk and bulge, and others appear extended or clumpy. As a result, a single slit cannot always cover a galaxy's full extent. However, as explained in Section~\ref{s:sed_fit}, we performed SED fitting on spectra and photometry extracted from within the slit region; we refer the reader to Bugiani et al. (in prep) for a detailed discussion on this matched-aperture SED fitting. 

We also use the morphological fits of these galaxies from NIRCam/F356W imaging, calculated in \citet{mahesonBigDustyGalaxies2025} with the \texttt{Pysersic} fitting tool \citep{pasha_pysersic_2023}. We refer the reader to that work for further information on the fitting procedure; however, it provided measurements of the axis ratio (b/a) for 21 of the 24 galaxies in the sample, which are explored in Section~\ref{s:comp_stellar}.

\begin{figure}
\centerline{\includegraphics[width=.5\textwidth]{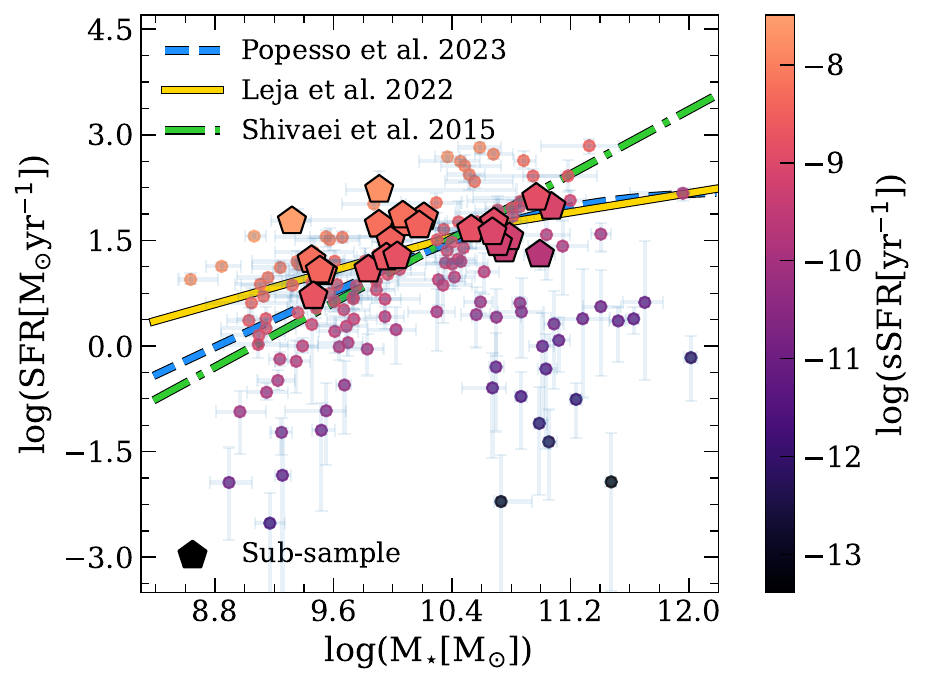}}
\medskip
\caption{The full Blue Jay sample of galaxies in the star-formation rate versus stellar mass plane, with the subsample of galaxies analysed in this work overplotted as pentagons. The star-forming main sequence (SFMS) presented in \citet{shivaeiMOSDEFSurveyDissecting2015}, \citet{leja_new_2022} and \citet{popessoMainSequenceStarforming2023} are overplotted as the blue dashed, green dashed-dot and yellow solid lines, respectively. The galaxies we select for in this work are along the SFMS, with some having very elevated SFRs. Although we preferentially select high-SFR galaxies through signal-to-noise cuts on the emission lines, we still have a wide range in stellar mass.} \label{f:SFMS}
\end{figure}

To compare this subsample with the full Blue Jay sample, we plot our galaxies in the stellar mass-SFR plane in Figure~\ref {f:SFMS}. We additionally overplot fits to the star-forming main sequence (SFMS) from \citet{shivaeiMOSDEFSurveyDissecting2015}, \citet{leja_new_2022} and \citet{popessoMainSequenceStarforming2023} for comparison. The star-forming galaxies in the full Blue Jay sample align well with the SFMS reported in the literature. Compared to the full sample, this subsample covers a similarly wide stellar-mass range. The galaxies in the subsample are mainly along the star-forming main sequence, with some being more star-forming, and cover a similarly wide redshift range ($1.8<z<3.3$) as the full sample ($1.6<z<3.7$).

We exclude all quiescent galaxies from the full Blue Jay sample because they lack strong hydrogen recombination emission lines. Thus, our selection criteria preferentially select highly star-forming galaxies with the strongest emission line fluxes, as well as galaxies with high dust attenuation, since the amount of attenuation is also strongly tied to the SFR \citep[e.g.,][]{nagaraj_bayesian_2022}. Although this biases our subsample relative to the full sample, we still cover a wide stellar-mass range. From previous studies, the stellar mass has been shown to be a strong driver of both the stellar and nebular optical dust attenuation \citep{garn_predicting_2010, maheson_unravelling_2024, mahesonBigDustyGalaxies2025} and the slope of the stellar continuum attenuation curve \citep{luNonParametricAttenuationCurves2025}.

\section{Measuring the Attenuation Curve}\label{s:calc_atten}

To measure the attenuation curve, we first need to calculate the predicted intrinsic line ratios to compare with our observed line ratios. In this work, we use the Python package \texttt{Pyneb} \citep{luridianaPyNebNewTool2015} to calculate these intrinsic line ratios using values of electron temperature ($T_e$) and electron density ($n_e$) for standard nebular conditions of $T_e = 10,000$~K and $n_e = 100$~cm$^{-3}$, as is done in other works such as \citet{reddyPaschenlineConstraintsDust2023}.

For some galaxies with reliable detections of auroral and strong emission lines, we can measure the electron temperature and density, though this was not the case for the full sample. However, since the Balmer emission lines are not produced through collisional excitation, the intrinsic emission line ratios are largely independent of the electron conditions over the typical ranges \citep[e.g.,][]{osterbrock_astrophysics_2006}. To show this, in Appendix~\ref{app:electron_t_n} we measure the electron temperature and density for the galaxies where possible and show that the intrinsic line ratios vary by less than $8\%$ across the range of electron conditions.

Using these standard electron conditions, we calculated the intrinsic line ratios relative to H${\alpha}$ and show them in Table~\ref{t:int_ratios}. We chose H${\alpha}$ as the reference line because it has the highest S/N in most of the sample, minimising the error propagated into the final dust attenuation curve. Any of the recombination lines can be used as a reference line, with \cite{sanders_aurora_2024} using the H${\beta}$ line and \cite{reddy_mosdef_2020} using H${\alpha}$. The choice of reference line does not alter the results; however, using lines with lower S/N than the H$\alpha$ emission line introduces unnecessary uncertainty. 

\begin{table}
  \centering
    \begin{tabular}{ccc}\hline
        Emission line &  $\lambda_{\rm rest}$ [\AA] & Ratio with H$\alpha$ \\ \hline
        H$\delta$ & 4102 & 0.090\\
        H$\gamma$ & 4340 & 0.164 \\
        H$\beta$ & 4861 & 0.349 \\
        H$\alpha$ & 6563 & 1.000 \\
        Pa9 & 9229 & 0.009 \\
        Pa$\delta$ & 10050 & 0.019 \\
        Pa$\gamma$ & 10939 & 0.032 \\
        Pa$\beta$  & 12819 & 0.057 \\ \hline
    \end{tabular}\\[5pt]
    \medskip
    \caption{Table showing the hydrogen recombination line ratios with the H$\alpha$ line assuming Case B recombination and $T_e=10,000$~K and $n_e=100$~cm$^{-3}$, calculated using \texttt{Pyneb}.}
    \label{t:int_ratios}
\end{table}

With these intrinsic line ratios, we can now calculate the attenuation of each line relative to the chosen reference line, H$\alpha$, to determine the nebular attenuation curve for each galaxy. In this work, we follow the methodology presented in \cite{reddy_mosdef_2020}, which we detail in this section. 

First, let us consider the attenuation of a single emission line at wavelength $\lambda$. The observed flux $f_o(\lambda)$ of the emission line is related to the intrinsic flux $f_i(\lambda)$ multiplied by a wavelength-dependent attenuation, or the attenuation curve, $A(\lambda)$.
\begin{align}
    f_o(\lambda) = &~ f_i(\lambda) \times 10^{-0.4A(\lambda)} \label{eq:atten_def}
\end{align}

If we then have two emission lines, at $\lambda_1$ and $\lambda_2$, the observed line ratio $f_o(\lambda_1)/f_o(\lambda_2)$ is related to the intrinsic line ratio $f_i(\lambda_1)/f_i(\lambda_2)$ through the following equation:
\begin{align}
    \frac{f_o(\lambda_1)}{f_o(\lambda_2)}=\frac{f_i(\lambda_1)}{f_i(\lambda_2)} \times 10^{-0.4[A(\lambda_1)-A(\lambda_2)]}
\end{align}
which can be rewritten as
\begin{align}
    A(\lambda_2) - A(\lambda_1) = &~ 2.5 \left[ \log \left(\frac{f_o(\lambda_1)}{f_0(\lambda_2)} \right) - \log \left(\frac{f_i(\lambda_1)}{f_i(\lambda_2)} \right) \right]  \label{eq:A_lam_1_2}
\end{align}

Therefore, we can measure the relative attenuation between the emission lines at $\lambda_1$ and $\lambda_2$ if we have the observed and intrinsic line ratio. 

If we want to measure the relative attenuation of various emission lines with respect to a fixed emission line at $\lambda_1$, we can measure the relative attenuation $A'(\lambda_2)$ using the following relation
\begin{align}
    A'(\lambda_2) ~ \equiv &~ A(\lambda_2)+[1-A(\lambda_1)] \label{eq:atten_shift} \\
    = &~ 2.5 \left[ \log \left(\frac{f_o(\lambda_1)}{f_0(\lambda_2)} \right) - \log \left(\frac{f_i(\lambda_1)}{f_i(\lambda_2)} \right) \right] + 1 \label{eq:atten_relative}
\end{align}
where we fix $\lambda_1$ to be the H$\alpha$ emission line in this work. Since we use several emission line ratios, with emission lines at different $\lambda_2$, we are probing the shape of the relative attenuation curve $A'(\lambda)$. This $A'(\lambda)$ is then offset from the absolute attenuation curve $A(\lambda)$ by a constant value ($1-A(\lambda_1)$) that is independent of wavelength, but will vary between galaxies. Therefore, we can measure the shape of $A'(\lambda)$ for each galaxy by using various emission line ratios (relative to H$\alpha$).

To better compare the attenuation curves we measure here with those from previous studies, we normalise the attenuation curve through the parameterisation $k(\lambda)$, as expressed in Equation~\ref{eq:k_lam_def},

\begin{align}
    k(\lambda) = &~ \frac{A(\lambda)}{E(B-V)_{\rm neb}} = \frac{A(\lambda)}{A(4400 \mathring{\rm A})-A(5500 \mathring{\rm A})} \label{eq:k_lam_def}
\end{align}
where $E(B-V)_{\rm neb}$ is the nebular reddening, i.e., the relative attenuation between the visible ($V$) band and the shorter-wavelength blue ($B$) band. Since the relative attenuation curve we measure in this work, $A'(\lambda)$, is offset from the absolute attenuation curve $A(\lambda)$ by a constant value, the reddening which is measured from the effective attenuation curve, $E'(B-V) = A'(4400\AA) - A'(5500\AA) = A(4400\AA) - A(5500\AA)$, is identical to the nebular reddening which would be measured from the absolute attenuation curve. Throughout the rest of this paper, we refer to the effective nebular reddening we measure as the nebular reddening.

Combining the parameterisation of the relative attenuation curve (Equation~\ref{eq:atten_shift}) and the definition in Equation~\ref{eq:k_lam_def}, we can express the relative attenuation curve $k'(\lambda)$ through:

\begin{align}\label{eq:k_prime}
    k'(\lambda) = &~ \frac{A'(\lambda)}{A'(4400 \mathring{\rm A})-A'(5500 \mathring{\rm A})} \\
    = & \frac{A(\lambda) + [1-A(\lambda_1)]}{E(B-V)_{\rm neb}}\nonumber \\
    = &~ k(\lambda) + C \quad , \nonumber \\
    & \text{where} \quad C =  \frac{1 - A(\lambda_1)}{E(B-V)_{\rm neb}} \nonumber
\end{align}

This relative attenuation curve $k'(\lambda)$ is thus offset from the absolute attenuation curve $k(\lambda)$ by a constant value independent of wavelength, which varies between galaxies. 

To measure the absolute attenuation curve from the relative attenuation curve, we need to determine the exact amount of attenuation at a certain wavelength to measure the normalisation constant $C$. This is commonly done by anchoring the attenuation curve at a longer wavelength, where the attenuation is roughly zero. In \citet{reddy_mosdef_2020} and \citet{reddyAURORASurveyMultiple2026a}, which use the same methodology to measure the nebular attenuation curves as in this paper, they anchor their attenuation curves by assuming the attenuation is equal to zero at a specific long wavelength ($2.8~\mu \rm m$) and extrapolating their attenuation curve. Since this extrapolation, as well as the choice of wavelength at which the attenuation is assumed to be zero, introduces large uncertainty in the normalisation, we choose to only consider the relative attenuation curve in this work. 

\section{Fitting the attenuation curve}\label{s:fitting_atten}

\begin{figure*}
\centerline{\includegraphics[width=\textwidth]{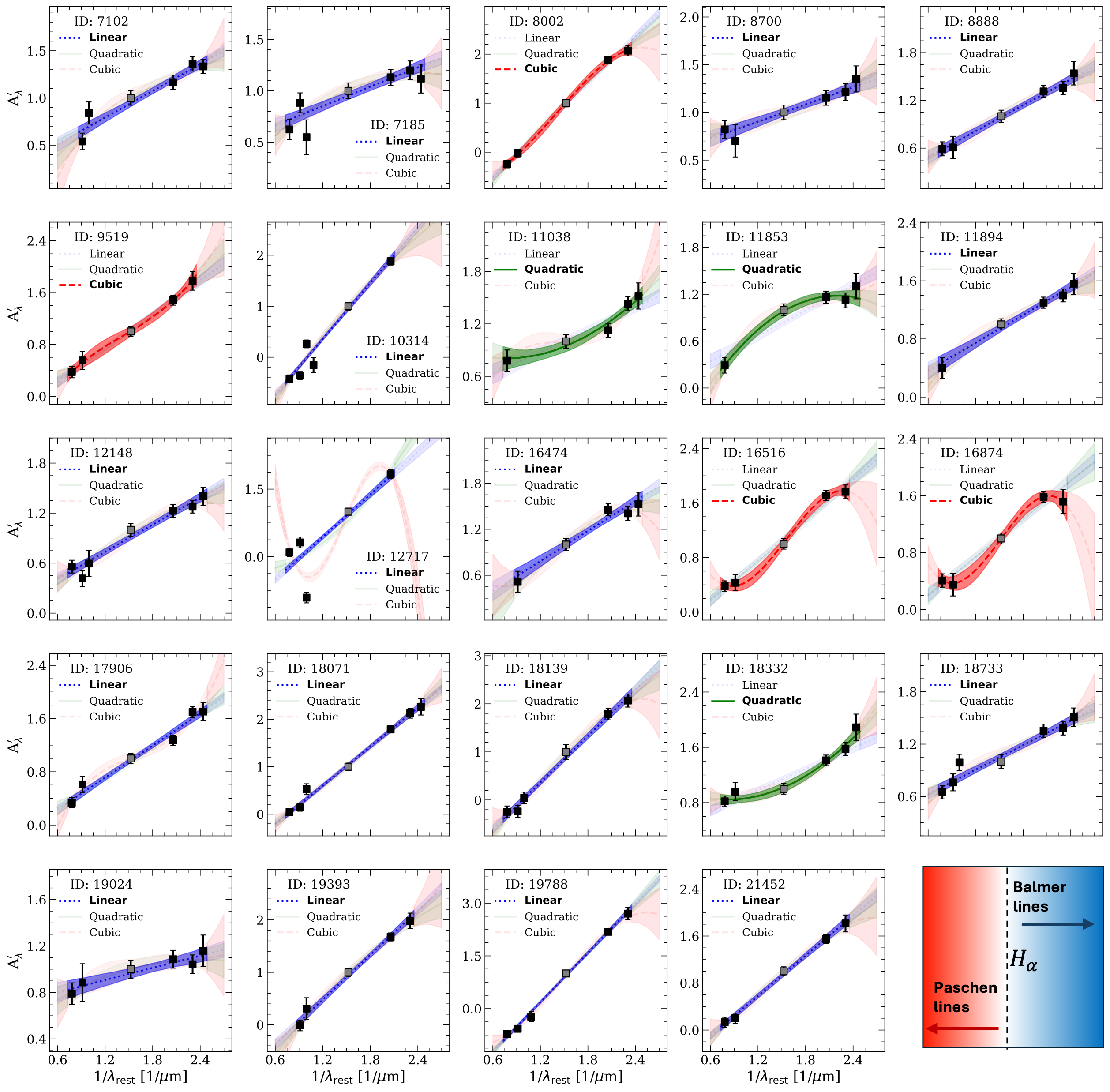}}
\medskip
\caption{The attenuation curve for each galaxy relative to the H$\alpha$ line plotted against the inverse of the rest-frame wavelength. The best-fit linear (blue dotted line), quadratic (green solid line) and cubic (red dashed line) relations are overplotted with their 16th-84th percentile range. The functional form which provides the best fit to the attenuation curve is indicated in bold in the legend for each galaxy. A cartoon showing where the Balmer and Paschen lines lie in each plot is shown in the bottom right panel for readability, and the H$\alpha$ data point is shown in grey on the plots.} \label{f:A_1_lam_grid}
\end{figure*}

Using the methodology outlined in the previous sections, we measure the relative attenuation curve at discrete wavelengths for each galaxy. These are presented in Figure~\ref{f:A_1_lam_grid} as the black squares.

Various functional forms are used in the literature to fit the attenuation curves. In \cite{calzetti_dust_2000}, they use a linear polynomial in 1/$\lambda$ to fit the rest-optical to NIR and a cubic polynomial for the rest-optical to UV. In \cite{cardelli_relationship_1989}, they use an 8th-order polynomial in the rest-optical to NIR and a power law of the form $1/\lambda^{1.61}$ for the rest-NIR to MIR to fit the MW extinction curve. We restrict the functional forms considered in this work to linear, quadratic, and cubic polynomials in $1/\lambda$ because each galaxy's attenuation curve has few data points. Similar functional forms are used in \cite{reddy_mosdef_2020} and \cite{sanders_aurora_2024}. 

We use the \texttt{EMCEE} fitting code \citep{foreman-mackey_emcee_2013} to fit each attenuation curve using the linear, quadratic, and cubic polynomials. This gave us the best fit of each polynomial (taken as the median of the posterior), along with the uncertainty on the best fit (taken as the 16th-84th percentile range). These attenuation curves are plotted in Figure~\ref{f:A_1_lam_grid} for each galaxy for each polynomial form.

For some galaxies, the choice of the functional form strongly affects the final shape of the attenuation curve due to over- or under-fitting. To determine which polynomial shape is best for each galaxy, we considered the following metrics: chi-squared, reduced chi-squared, Bayesian information criterion \citep[BIC;][]{neathBayesianInformationCriterion2012} and Akaike information criterion \citep[AIC;][]{cavanaughAkaikeInformationCriterion2019}. The chi-squared value (the sum of the residuals ([fit-truth]/error)$^2$) is commonly used to select the best model; however, here, the chi-squared always selects the cubic fit as the best fit. Instead, it is better to use a metric that accounts for the number of data points and fit parameters, such as reduced chi-squared, BIC, and AIC, to avoid overfitting. Each of these metrics penalises increasing the number of fit parameters with respect to the number of data points \citep{stoicaModelorderSelectionReview2004}. These three metrics all agree on which functional form best fits the attenuation curve for each galaxy. 

The preferred fit is indicated in bold for each galaxy in Figure~\ref{f:A_1_lam_grid}, and the coefficients of the polynomial fit for each galaxy are shown in Table~\ref{t:fit_params}. Overall, 17 galaxies prefer a linear fit, 3 prefer a quadratic fit, and 4 prefer a cubic fit.

\begin{table*}
  \centering
    \begin{tabular}{cccccc}\hline
        ID & Fit & $c_0$ & $c_1$ & $c_2$ & $c_3$  \\\hline
        7102 & Linear & $0.23 \pm 0.11$ & $0.47 \pm 0.06$ & $-$ & $-$ \\
    7185 & Linear & $0.47 \pm 0.10$ & $0.31 \pm 0.06$ & $-$ & $-$ \\
    8002 & Cubic & $-0.24 \pm 1.52$ & $-1.49 \pm 3.41$ & $2.36 \pm 2.34$ & $-0.55 \pm 0.50$ \\
    8700 & Linear & $0.55 \pm 0.12$ & $0.30 \pm 0.07$ & $-$ & $-$ \\
    8888 & Linear & $0.16 \pm 0.12$ & $0.54 \pm 0.06$ & $-$ & $-$ \\
    9519 & Cubic & $-1.01 \pm 1.74$ & $2.67 \pm 3.92$ & $-1.38 \pm 2.73$ & $0.33 \pm 0.59$ \\
    10314 & Linear & $-1.83 \pm 0.09$ & $1.83 \pm 0.07$ & $-$ & $-$ \\
    11038 & Quadratic & $0.94 \pm 0.42$ & $-0.36 \pm 0.55$ & $0.24 \pm 0.16$ & $-$ \\
    11853 & Quadratic & $-0.95 \pm 0.36$ & $1.94 \pm 0.50$ & $-0.44 \pm 0.16$ & $-$ \\
    11894 & Linear & $-0.02 \pm 0.17$ & $0.64 \pm 0.09$ & $-$ & $-$ \\
    12148 & Linear & $0.08 \pm 0.09$ & $0.54 \pm 0.06$ & $-$ & $-$ \\
    12717 & Linear & $-1.45 \pm 0.13$ & $1.57 \pm 0.10$ & $-$ & $-$ \\
    16474 & Linear & $0.00 \pm 0.18$ & $0.65 \pm 0.09$ & $-$ & $-$ \\
    16516 & Cubic & $2.69 \pm 1.52$ & $-6.11 \pm 3.50$ & $4.86 \pm 2.46$ & $-1.04 \pm 0.53$ \\
    16874 & Cubic & $3.38 \pm 1.97$ & $-7.77 \pm 4.51$ & $6.13 \pm 3.17$ & $-1.35 \pm 0.69$ \\
    17906 & Linear & $-0.25 \pm 0.10$ & $0.80 \pm 0.06$ & $-$ & $-$ \\
    18071 & Linear & $-1.03 \pm 0.09$ & $1.36 \pm 0.06$ & $-$ & $-$ \\
    18139 & Linear & $-1.57 \pm 0.13$ & $1.61 \pm 0.09$ & $-$ & $-$ \\
    18332 & Quadratic & $1.09 \pm 0.32$ & $-0.59 \pm 0.48$ & $0.36 \pm 0.15$ & $-$ \\
    18733 & Linear & $0.35 \pm 0.08$ & $0.47 \pm 0.05$ & $-$ & $-$ \\
    19024 & Linear & $0.70 \pm 0.12$ & $0.18 \pm 0.07$ & $-$ & $-$ \\
    19393 & Linear & $-1.23 \pm 0.16$ & $1.42 \pm 0.09$ & $-$ & $-$ \\
    19788 & Linear & $-2.63 \pm 0.10$ & $2.34 \pm 0.07$ & $-$ & $-$ \\
    21452 & Linear & $-0.78 \pm 0.10$ & $1.14 \pm 0.07$ & $-$ & $-$ \\
    
    \hline
    \end{tabular}\\[5pt]
    \medskip
    \caption{Best-fit parameters for the nebular attenuation curve for each galaxy of the form $A'(\lambda) = c_0 + c_1/\lambda + c_2/\lambda^2 + c_3/\lambda^3$ for $\lambda$ in $\mu$m.}
    \label{t:fit_params}
\end{table*}

\subsection{Nebular Reddening from Balmer and Paschen lines}\label{s:obsc_sf}

From this measured relative attenuation curve $A'(\lambda)$, we are able to directly measure the nebular reddening, $E(B-V)$, by taking $A'(B) - A'(V)$, where $B$ and $V$ denote the wavelengths of the B and V bands, at $4400\AA$ and $5500\AA$ respectively. However, for some galaxies, such as 11853, 16516 and 16874 in particular, the attenuation curve appears to "flatten" around the B and V bands, which is around $2.0 \rm \mu m^{-1}$ in Figure~\ref{f:A_1_lam_grid}, or where H$\delta$ and H$\gamma$ reside. This makes the measured reddening very small and consistent with zero. For those same galaxies, the attenuation curve is much steeper at redder wavelengths where the Paschen lines lie, and if we considered only the Paschen lines, the inferred reddening may be larger. 

This scenario, in which the reddening inferred from the Balmer lines is less than that from the Paschen lines, could be caused by obscured star formation. In this case, the dust column density toward certain star-forming regions is high enough to completely attenuate some Balmer lines, although it is optically thin to the Paschen lines. Therefore, looking only through the Balmer lines may probe only the least dusty regions where the Balmer lines can escape the galaxy, implying lower reddening than is true. If so, measuring reddening through the Paschen lines would be more accurate. The result that Paschen lines imply larger reddening than Balmer lines has been reported in \citet{prescottUsingMultipleEmission2022}, \citet{gimenez-arteagaSpatiallyResolvedProperties2023}, \citet{reddyPaschenlineConstraintsDust2023} and \citet{reddyAURORASurveyMultiple2026a}.

\begin{figure}
\centerline{\includegraphics[width=.5\textwidth]{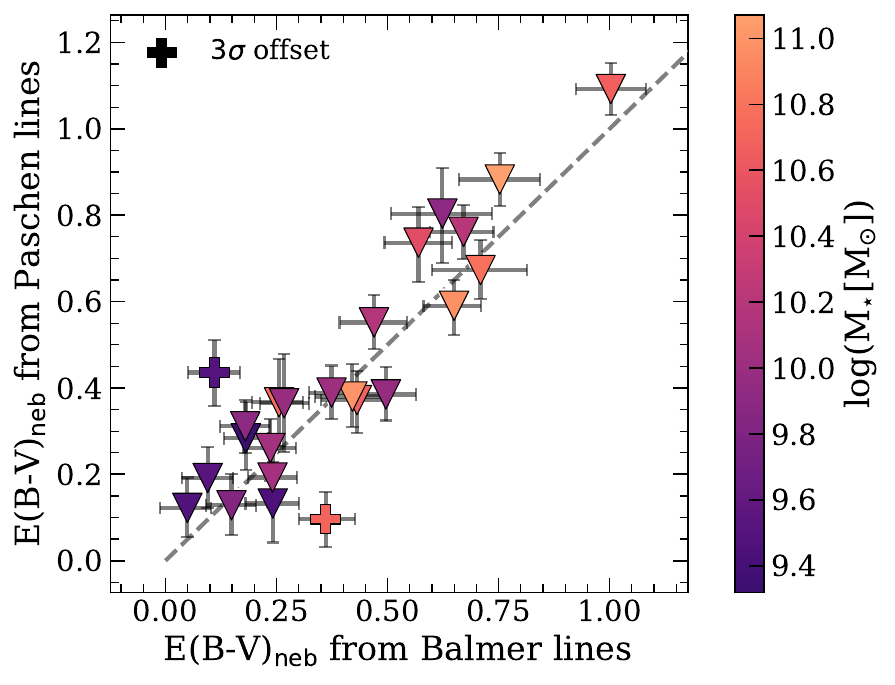}}
\medskip
\caption{Nebular reddening derived from the Paschen lines plotted against nebular reddening from the Balmer lines, colour-coded by stellar mass. Most galaxies have consistent tracers of the nebular reddening, aligning with the 1:1 line (grey dashed-dot). One galaxy has Paschen reddening at least $3 \sigma$ greater than the 1:1 line, and one galaxy has Balmer reddening at least $3 \sigma$ below the 1:1 line, denoted by a cross symbol instead of an inverted triangle.} \label{f:E_BV_B_P}
\end{figure}

To quantify this effect and test whether it is significant in our galaxies, we first measured the reddening through only the Paschen lines and then only the Balmer lines. We repeat the methodology presented in Section~\ref{s:calc_atten}, first using the ratio of only the Paschen lines with H$\alpha$ and then only the Balmer lines with H$\alpha$, and fitting only a linear relation in $1/\lambda$ because of the smaller number of data points. We then use this fit to measure the reddening of the Balmer and Paschen lines, which we present in Figure~\ref{f:E_BV_B_P}. The galaxies are distributed around the 1:1 line, showing no significant trend toward larger reddening from the Paschen lines. Of the sample considered here, only two galaxies are significantly ($>3\sigma$) offset from the 1:1 line (shown as crosses), meaning that for more than 90\% of the sample, the Balmer and Paschen inferred reddening are equivalent, and thus obscured star formation is not substantially affecting the sample as a whole.

The galaxies that show a significant offset between their Balmer- and Paschen-derived reddening are 11853 (greater Paschen than Balmer reddening) and 18332 (greater Balmer than Paschen reddening). The elevated Paschen reddening relative to Balmer in 11853 may imply some obscured star formation, so the Balmer reddening underpredicts the total dust attenuation. The elevated Balmer reddening of 18332, however, cannot be explained by obscured star formation, but could potentially be caused by non Case-B recombination.

Another galaxy, 16874, does not have a significant offset between the Balmer- and Paschen-derived reddening even though the shape of the attenuation curve clearly "flattens" around the Balmer lines. This is because the attenuation curve also flattens around the Paschen lines. If we instead use a linear fit on only the Paschen lines, the overall slope of the curve will be captured, and we can then measure the true nebular reddening for this galaxy. 

\section{Attenuation Curve Comparisons}\label{s:comp_studies}

\begin{figure*}
\centerline{\includegraphics[width=\textwidth]{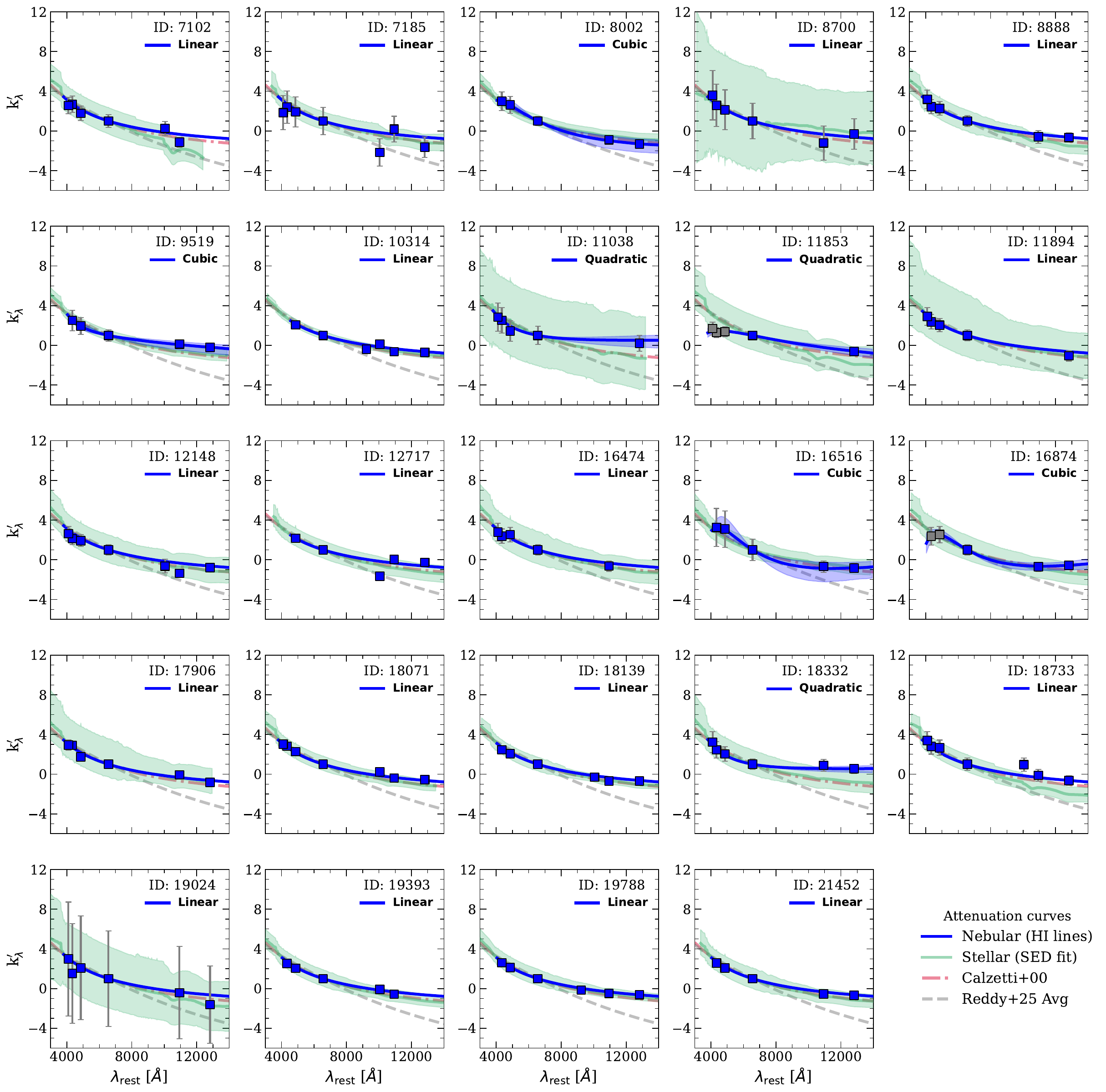}}
\medskip
\caption{The attenuation curve for each galaxy, $k'(\lambda)$, shifted to be zero at H$\alpha$ ($6563$\AA), plotted against the rest-frame wavelength as the blue curve and data points. The preferred form of the attenuation curve (linear, quadratic, or cubic) is used for each galaxy and is indicated on the plots. The stellar continuum attenuation curves derived through SED fitting are shown as the green curve. The \citealt{calzetti_dust_2000} attenuation curve is shown as the red dashed-dot line, and the average nebular attenuation curve derived in \citet{reddyAURORASurveyMultiple2026a} using the same methodology as in this work is shown as the grey dashed line. Our nebular attenuation curves show little shape diversity, and most galaxies align well with the slope of the \citet{calzetti_dust_2000} attenuation curve. For some galaxies, the stellar continuum attenuation curve has large error bars, but for every galaxy, the nebular and stellar attenuation curves overlap.} \label{f:k_lam_grid}
\end{figure*}

In the previous section, we determined the preferred shape of the attenuation curve for each galaxy. Here we will use this to determine the normalised relative attenuation curve $k'(\lambda)$ (Equation~\ref{eq:k_prime}) for each galaxy to compare with other attenuation curves in the literature. The nebular reddening $\rm E(B-V)_{\rm neb}$ used to calculate $k'(\lambda)$ was measured from the best-fit attenuation curve for each galaxy, except for two galaxies (11853 and 16874). For these, we used only the Paschen lines to determine the reddening used to obtain $k'(\lambda)$, since the reddening derived from the full shape of the curve did not well represent the reddening of the galaxy, as described in the previous section. 

We plot $k'(\lambda)$ in Figure~\ref{f:k_lam_grid} along with several curves from the literature, all shifted such that $k'(\rm H\alpha)=1$. The Balmer lines not used to derive the reddening for galaxies with ID 11853 and 16874 are shown as grey points in Figure~\ref{f:k_lam_grid}. Overall, we find little diversity in the shapes of our nebular attenuation curves, and most align with the \citet{calzetti_dust_2000} curve.

One may notice that the slope for all the curves which prefer a linear fit of $A'(\lambda)$ in 1/$\lambda$ seems very similar. This is because the slope is in fact fixed to a constant value. This is clear if we consider the functional form for the linear fit and use the fact that $E(B-V) = A(B) - A(V)$: 
\begin{align}
    A'(\lambda) =& a + b/\lambda, \text{ then } E'(B-V) = b(1/4500\mathring{A}  -1/5500\mathring{A}), \\ \nonumber
\end{align}

From Equation~\ref{eq:k_lam_def}, $k'(\lambda)=A'(\lambda)/E(B-V)$, and we can thus express $k'(\lambda)$ for a galaxy with a linear fit of $A'(\lambda)$ in 1/$\lambda$ as:

\begin{align}
    k'(\lambda)= &a/b*1/(1/4500\mathring{A} -1/5500\mathring{A}) \\
    &+1/\lambda*1/(1/4500\mathring{A} -1/5500\mathring{A})
\end{align}

Therefore, the slopes of the normalised attenuation curves are identical for all galaxies which prefer a linear fit of $A'(\lambda)$ in 1/$\lambda$. Since this applies to most galaxies in our sample, the resulting normalised attenuation curves show little shape diversity. Their slopes also closely match those of the \citet{calzetti_dust_2000} attenuation curve, as they are nearly coincident in Figure~\ref{f:k_lam_grid}. Even for the galaxies which prefer a quadratic or cubic fit, and thus have a slope in $k'(\lambda)$ which can vary, the curves are not significantly different to the \citet{calzetti_dust_2000} curve. This is in stark contrast with the results from \citet{reddyAURORASurveyMultiple2026a}, where they find strong diversity and an average curve significantly steeper than the \citet{calzetti_dust_2000} curve. 

The difference between the attenuation curves measured in our two studies is surprising, since we both study emission-line-bright star-forming galaxies at cosmic noon. In the next section, we explore how our two samples differ and what may be causing the difference in attenuation-curve shapes.  

\subsection{Comparison with \citet{reddyAURORASurveyMultiple2026a}}

\begin{figure}
\centerline{\includegraphics[width=.5\textwidth]{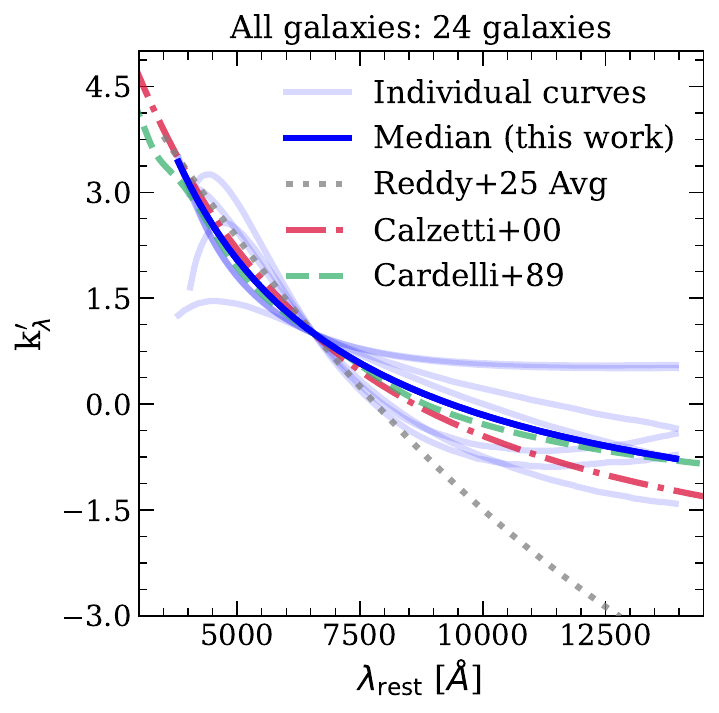}}
\medskip
\caption{Our median nebular attenuation curve ($k'(\lambda)$, in dark blue), plotted next to the individual galaxy attenuation curves (in pale blue), along with the average nebular attenuation curve from \citet{reddyAURORASurveyMultiple2026a} as the grey dotted line, the \citet{calzetti_dust_2000} attenuation curve for local starbursts as the red dashed-dot line, and the \citet{cardelli_relationship_1989} Milky Way extinction curve as the green dashed line. All curves plotted are shifted such that the attenuation at H$\alpha$ is unity. Since most galaxies prefer a linear fit of $A'(\lambda)$ in 1/$\lambda$, which leads to identical $k'(\lambda)$ curves (as described in the text), the median curve overlaps identically with the shape of the linear fit, which is similar in shape to the Milky Way extinction curve and slightly steeper than the \citet{calzetti_dust_2000} curve in this wavelength regime. Out of the sample of 24 galaxies, only 6 show any divergence from this linear relation, and none are as steep as the average curve from \citet{reddyAURORASurveyMultiple2026a}.} \label{f:k_lam_median}
\end{figure}

To show that our attenuation curves have minimal diversity and that they are all significantly shallower than the average \citet{reddyAURORASurveyMultiple2026a} attenuation curve, we plot the individual curves, along with the median, in Figure~\ref{f:k_lam_median}. Since the majority of the galaxies prefer a linear fit of $A'(\lambda)$ in 1/$\lambda$, the median attenuation curve aligns exactly with the shape of the linear fit, which is $\left< k'(\lambda)_{\rm neb}\right> = -2.35 + 2.20/(\lambda ~ \rm [\mu m])$. From Figure~\ref{f:k_lam_median}, our median curve is significantly shallower than the \citet{reddyAURORASurveyMultiple2026a} curve, aligns well with the Milky Way extinction curve, and is slightly shallower than the \citet{calzetti_dust_2000} curve. The curves also show little diversity in shape, as most prefer a linear fit, and no individual attenuation curve is as steep as the average from \citet{reddyAURORASurveyMultiple2026a}. 

\begin{figure}
\centerline{\includegraphics[width=.5\textwidth]{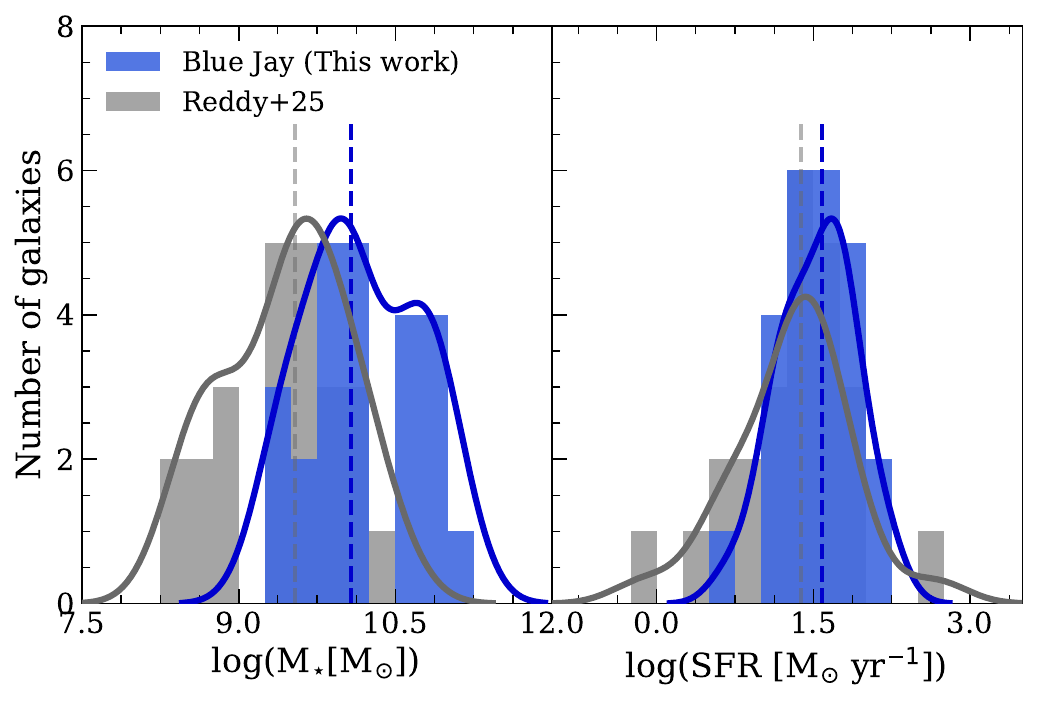}}
\medskip
\caption{Comparison of the stellar mass and SFR distribution for the galaxies in this sample (Blue Jay, in blue) compared to that from \citet{reddyAURORASurveyMultiple2026a} (in grey). The Blue Jay galaxies are shifted to higher masses than in the other survey, with a difference in median mass of 0.5 dex, whilst the SFR distributions more closely overlap, with a difference in medians of 0.2 dex.} \label{f:sample_hist}
\end{figure}

These results, compared with those from \citet{reddyAURORASurveyMultiple2026a}, raise a question: why do the nebular attenuation curves differ so much between our two samples? There are several potential reasons for this difference, such as different sample selection techniques, which can lead to different physical properties of the galaxies, or differences in the data reduction techniques, such as the slit loss corrections. 

The most likely reason for differences in our attenuation curves is sample selection. The Blue Jay survey was selected to be representative of the entire galaxy population at these redshifts, without strong biases in star formation rate or other properties \citep[see][for further information]{belliBlueJaySurvey2025}. The AURORA survey, however, selected galaxies which had a high probability of having bright auroral lines, likely biasing its sample toward high-electron-temperature, low-metallicity galaxies \citep{shapleyAURORASurveyNew2025}. These biases may then lead to differences in the dust properties of the galaxies they observe compared with our sample.

To compare the stellar population properties of our two samples, we plot the stellar mass and SFR distributions in Figure~\ref{f:sample_hist}. The stellar mass distributions differ clearly, with our sample extending to higher masses and a median 0.5 dex higher, whereas the SFR distributions are similar. However, the difference in the stellar mass distributions is unlikely to be the sole cause of the difference in the attenuation curves, since there is still substantial overlap in stellar mass, yet the attenuation curves show almost no overlap in shape. To emphasise this, in Appendix~\ref{app:med_atten_comp} we measure the median nebular attenuation curve for our low-mass ($\log(\rm M_{\star}~[M_\odot])<10$) and then high-mass ($\log(\rm M_{\star}~[M_\odot])>10$) galaxies, and we see no variation in the level of diversity between the two samples.

Additionally, our redshift ranges differ: we probe a slightly lower range ($1.8<z<3.3$) than the \citet{reddyAURORASurveyMultiple2026a} sample ($1.5<z<4.4$). In Appendix~\ref{app:med_atten_comp}, we also measure the median nebular attenuation curve for our high ($z\geq2.5$) and then low ($z<2.5)$ redshift galaxies, and the high redshift galaxies are shown to be the most diverse in the sample. This could potentially be the cause of the differences in attenuation curve slopes we see, with the nebular attenuation curves steepening and becoming more diverse at redshifts beyond what we probe in this work. However, most evidence studying stellar attenuation curves \citep[e.g.,][]{markovUnveilingTrendsDust2025a, shivaeiDiversityEvolutionDust2025} points to a steady shallowing of slope as redshift increases, although it is not clear whether stellar and nebular attenuation curves will vary with redshift in the same way. Further studies across a wider redshift range are needed to clarify this. 

Another potential driver of differences between our results may be due to variations in Balmer- and Paschen-derived reddening. In \cite{reddyAURORASurveyMultiple2026a}, they find that many of their galaxies show elevated Paschen reddening relative to Balmer, and associate this with varying star-dust geometry, which they argue causes the diversity in their nebular attenuation curves. However, we find little difference between the Balmer- and Paschen-derived reddening (see Section~\ref{s:obsc_sf}), suggesting that the star-dust geometry does not vary significantly between our galaxies and may explain the lack of diversity in our nebular attenuation curves.

Beyond the differences being physically motivated, they may instead reflect differences in data reduction or calibration methods. For example, how the slit loss corrections are handled; since we are dealing with line ratios across a very broad wavelength range, any difference in the wavelength-dependent slit-loss corrections will significantly affect the measured curves.

With only 24 galaxies in each study, we cannot determine exactly what drives the differences between the nebular attenuation curves we measure. To better address this problem, we require further studies observing more individual galaxies across larger redshift ranges to explore what is driving or suppressing diversity. Until then, we cannot definitively say whether other studies should dust-correct their emission lines using the nebular attenuation curve derived here or the one presented in \citet{reddyAURORASurveyMultiple2026a}.

\subsection{Comparison with the Stellar Continuum attenuation curve}

Now turning our attention to the stellar continuum attenuation curves of these galaxies, which are shown as the pale green curves in Figure~\ref{f:k_lam_grid}, we see that either they agree well with the nebular attenuation curve or their error regions are so large that the curves are nearly unconstrained. 

For the galaxies with well-constrained stellar attenuation curves, the curves generally align with the nebular attenuation curve. Although the stellar dust model employs a modified \citet{calzetti_dust_2000} curve (see Section~\ref{s:sed_fit}), the slope of the stellar attenuation curve is still free, as is clear in e.g. 7102 where the stellar curve is as steep as the \citet{reddyAURORASurveyMultiple2026a} average curve, albeit with large error bars. The coincidence of many stellar and nebular attenuation curves could be physically motivated; since the sSFR of the galaxies in this sample is very high, young stars likely dominate the spectrum, so the nebular and stellar light undergo similar attenuation. Additionally, these galaxies have a median stellar $\rm A_V$ of $~1.0$, and from both \citet{mahesonBigDustyGalaxies2025} (which studied the stellar continuum attenuation curves of the entire Blue Jay sample) and \citet{shivaeiDiversityEvolutionDust2025}, the attenuation curves flatten and approach the slope of the \citet{calzetti_dust_2000} curve at high attenuation ($\gtrsim1$).

The large error regions in some stellar attenuation curves stem from broad posteriors on the dust parameters, likely because the galaxies are very faint; 7102, 11038, 11853 and 19024 have the lowest F444W flux in the sample. Or, since metallicity and dust can be coupled, uncertainty in the dust parameters may stem from poorly constrained metallicity, as in 8700, 11894 and 17906, which have few absorption lines and thus have poorly constrained metallicities. 

Additionally, one may notice the bump around the Paschen lines in the stellar continuum attenuation curve. This is not the Paschen break, but a PAH dust emission feature. This appears in our stellar continuum attenuation curve because of how we measure this for each galaxy. As described in Section~\ref{s:sed_fit}, we take the model spectra with dust turned on and off to obtain the total stellar continuum attenuation curve. As a consequence of the dust being turned off, both attenuation and emission are turned off, and thus the dust emission features will show up in this curve.

Not only are we able to extract the stellar continuum attenuation curve from the SED fit, but we are also able to obtain the nebular attenuation curve since the two-component dust model is employed. In Appendix~\ref{app:neb_sed}, we show that the nebular reddening measured through emission lines agrees overall with the SED-derived nebular reddening.

\section{Nebular and stellar reddening}\label{s:comp_stellar}

\begin{figure}
\centerline{\includegraphics[width=.5\textwidth]{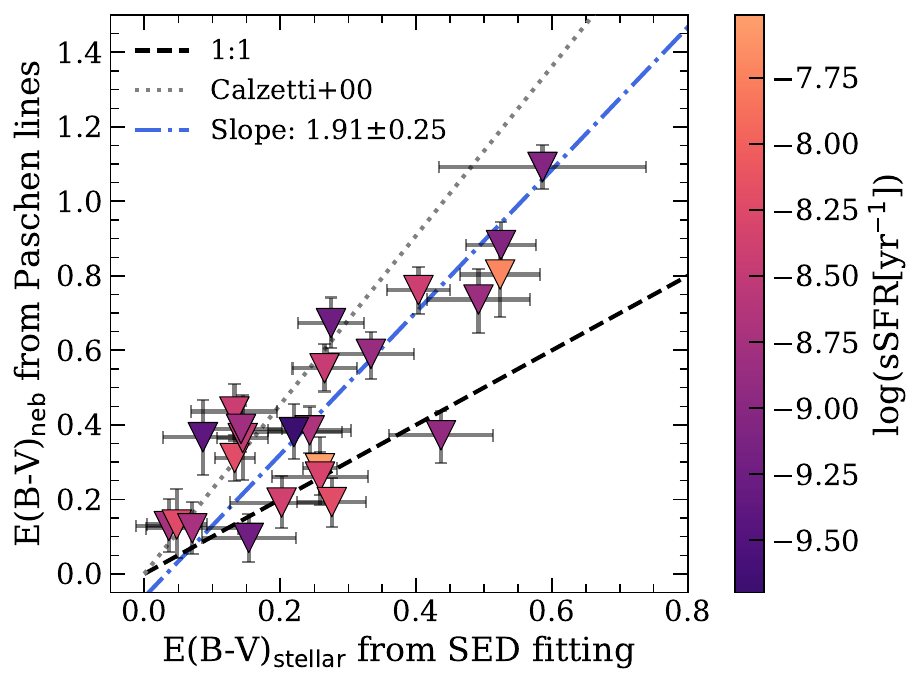}}
\medskip
\caption{The nebular reddening, E(B-V)$_{\rm neb}$, measured from the Paschen lines plotted against the stellar reddening, E(B-V)$_{\rm star}$, measured from SED fitting, colour coded by the sSFR. Our galaxies are distributed between the 1:1 line (black dashed line) and the roughly 2.27:1 line from \citet{calzetti_dust_2000} (grey dotted line). The line of best fit to our measurements (blue dot-dashed line) shows that the nebular reddening is on average 1.9 times larger than the stellar reddening.} \label{f:reddening_neb_stellar_SED}
\end{figure}

In this section, we explore the relationship between nebular reddening derived from the Paschen lines and properties derived from SED-fitting, such as stellar reddening, stellar mass, and SFR. Throughout this section, we use the Paschen-derived nebular reddening because it is less affected by obscured star formation and better traces the overall reddening for some galaxies in this sample, as discussed in Section~\ref{s:obsc_sf}.

The difference between nebular and stellar reddening has long been studied, and what is driving variations in this diference has been attributed to the sSFR \citep{price_direct_2014, puglisi_dust_2016}, the SFR \citep{reddy_mosdef_2015, lorenzStackingAnalyzingMOSDEF2024}, the metallicity \citep{shivaei_mosdef_2020}, the stellar mass \citep{lorenzStackingAnalyzingMOSDEF2024} and the inclination \citep{battistiCharacterizingDustAttenuation2017}. However, when these studies compare stellar and nebular reddening, many carry a systematic uncertainty in the choice of the nebular attenuation curve: a steeper attenuation curve yields smaller nebular reddening for a fixed hydrogen recombination line ratio, affecting the comparison between nebular and stellar reddening. In this work, we use the nebular reddening associated with the nebular attenuation curve measured directly from hydrogen recombination lines, removing this uncertainty.

We plot the nebular reddening from the Paschen lines against the stellar reddening from SED fitting in Figure~\ref{f:reddening_neb_stellar_SED}, and we see that they are positively correlated, with the data points scattered between the 1:1 line and the canonical \citet{calzetti_dust_2000} relation of 2.27:1. For our galaxies, there is a large spread in how the nebular and stellar attenuation are related to each other, and the line of best fit to our data shows that on average the nebular reddening is a factor of $1.91$ greater than the stellar reddening, $1~\sigma$ less the value from \citet{calzetti_dust_2000}, however still consistent with the nebular reddening being elevated on average compared to the stellar reddening.

To further explore how the nebular and stellar reddening relate to each other and other galaxy properties, we measured the differential reddening, defined as $\Delta \rm E(B-V) \equiv E(B-V)_{neb} - E(B-V)_{stellar}$. Recently, \citet{karthikeyanBalmerDecrementsNebularStellar2026} found that at redshifts $z<4$, more massive and more star-forming galaxies show greater differential reddening than lower-mass or less star-forming galaxies. However, at higher redshifts ($4<z<7$), they find no variation in differential reddening, implying that stellar continuum and nebular light probe similar dust columns. This trend at $z<4$ has also been observed by other works at similar redshift ranges \citep[e.g.,][]{reddy_mosdef_2015, sandersMOSDEFSurveyEvolution2021, lorenz_updated_2023}.

\begin{figure}
\centerline{\includegraphics[width=.5\textwidth]{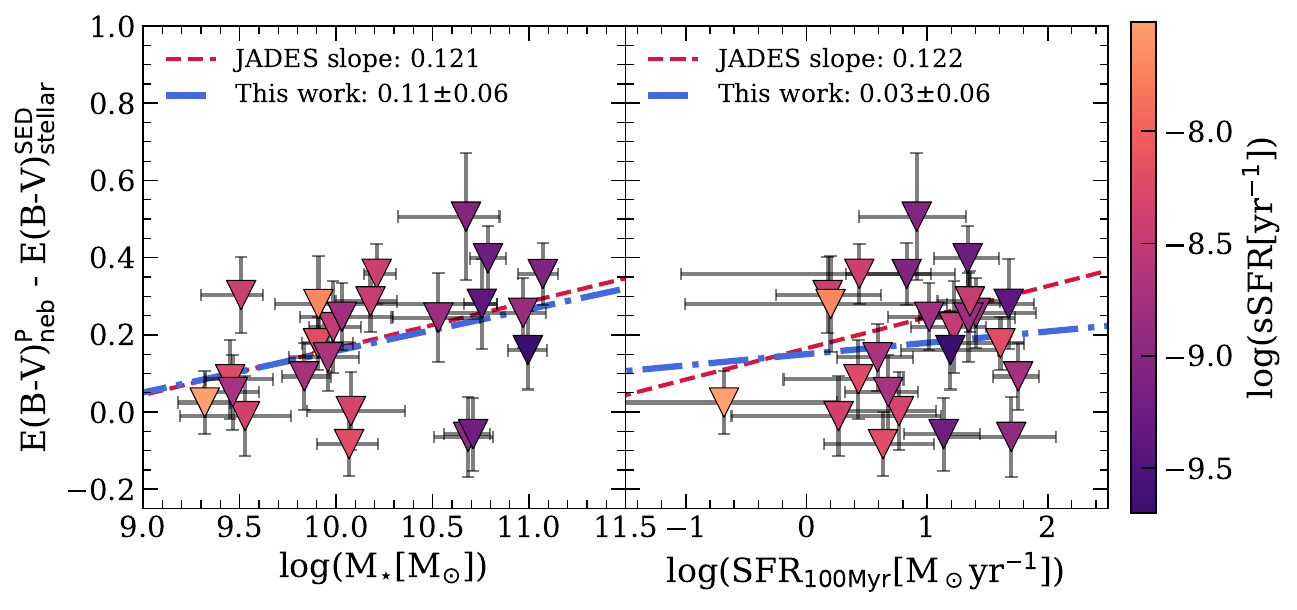}}
\medskip
\caption{The differential reddening (E(B-V)$_{\rm neb}$ from the Paschen lines minus E(B-V)$_{\rm stellar}$ from SED fitting) plotted against the stellar mass and SFR (measured over the last 100 Myr), colour-coded by the sSFR. We include the line of best fit for these galaxies and the relation measured from the JADES survey \citep{karthikeyanBalmerDecrementsNebularStellar2026}. More massive galaxies show a weak trend toward greater differential reddening, and our lines of best fit agree within $1~\sigma$. There is no such trend between the SFR and differential reddening, with our slope consistent with zero and significantly shallower than the literature relation.} \label{f:red_neb_SED_diff}
\end{figure}

This work is well placed to address whether this trend is present at low redshifts and if the choice of nebular attenuation curve plays a role. We plot our differential reddening against the stellar mass and then the SFR in Figure~\ref{f:red_neb_SED_diff}. In the left panel, a very weak correlation can be seen with stellar mass: more massive galaxies, on average, appear to have greater differential reddening than lower-mass galaxies. This is not the case in the right panel, where differential reddening shows no trend with SFR. We also include relationships between differential reddening and stellar mass and SFR from \citet{karthikeyanBalmerDecrementsNebularStellar2026} for their lowest-redshift bin, which covers $2.5<z<4$, thus extending to higher redshifts than our sample. For comparison, we include the line of best fit to our data points in Figure~\ref{f:red_neb_SED_diff}, and we can see that in the left panel with stellar mass, our line of best fit aligns within $1~\sigma$ of the line of best fit from \citet{karthikeyanBalmerDecrementsNebularStellar2026}; however, with the SFR, our line of best fit is significantly shallower. Since the relation with SFR reported in \citet{karthikeyanBalmerDecrementsNebularStellar2026} uses the SFR averaged over the last $100$~Myr, we likewise use the SFR of these galaxies measured over the last $100$~Myr in this section for consistency. 

These results are in partial agreement with those presented in \citet{lorenz_updated_2023}, where they identify an increase in the differential reddening with stellar mass and SFR for stacked spectra of galaxies at $1.3<z<2.6$ (i.e., the nebular reddening exceeds the stellar reddening more so at high masses than at low masses), as well as a lack of inclination dependence on either the stellar or nebular reddening. The standard two-component dust model from \citet{charlot_simple_2000} can not explain either of these results, and as such, a three-component dust model is put forward in \citet{lorenz_updated_2023}. In the standard two-component dust model, differential reddening is not predicted to correlate with SFR or stellar mass alone, and stellar and nebular reddening should be somewhat inclination-dependent due to attenuation by diffuse ISM dust. This three-component dust model proposed by \citet{lorenz_updated_2023} is equivalent to the two-component dust model, with the added contribution of massive, dusty star-forming clumps that contain mostly young stars. These clumps have been observed to be larger \citep{swinbankPROPERTIESSTARFORMINGINTERStelLAR2012} and more common \citep{tadakiNATUREHaSELECTEDGALAXIES2013} as the stellar mass increases. Because these clumps mainly contain young stars, they contribute most to the nebular emission, while those young stars and stars outside the clumps contribute to the stellar continuum. As stellar mass increases, these clumps become more common, leading to the nebular emission being significantly more attenuated than the stellar continuum on average. 

\begin{figure}
\centerline{\includegraphics[width=.5\textwidth]{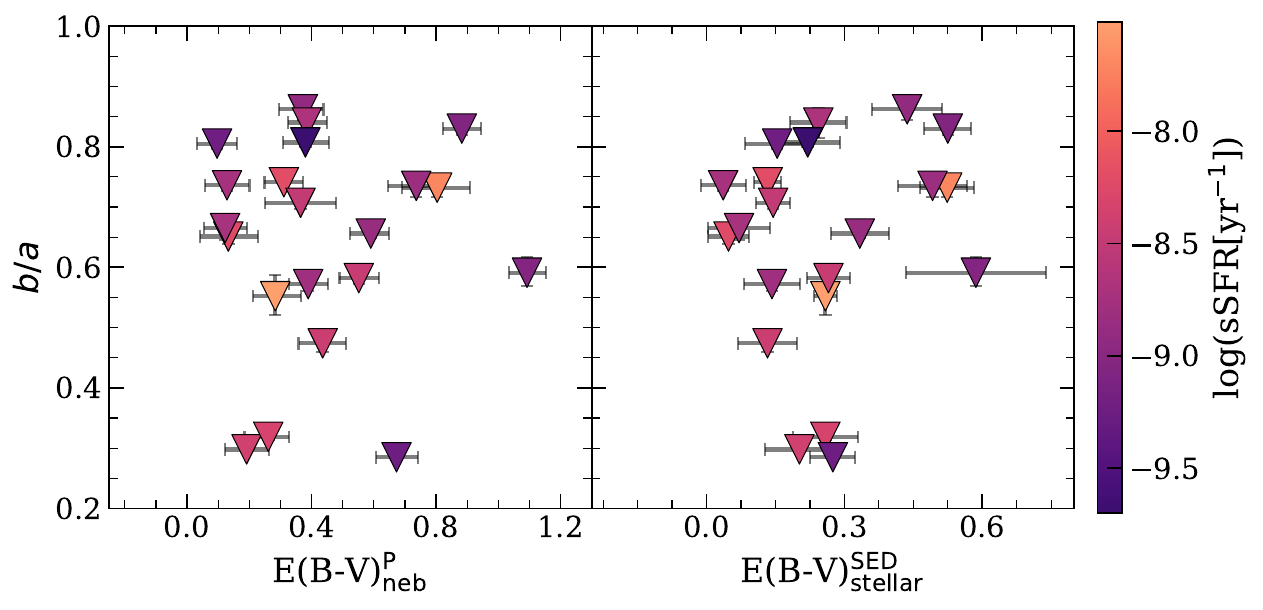}}
\medskip
\caption{The axis ratio (b/a) plotted against the nebular reddening (E(B-V)$_{\rm neb}$ from the Paschen lines) in the left panel, and the stellar reddening (E(B-V)$_{\rm stellar}$ from SED fitting) in the right panel, colour-coded by the sSFR over the last 30 Myr. The axis-ratio errors are plotted, although they are too small to see by eye. In both panels, there is no significant correlation between the reddening and axis ratio, implying that the dust distribution dominating the attenuation of the nebular and stellar light is roughly spherically symmetric.} \label{f:b_a_red}
\end{figure}

Additionally, in this three-component dust model, most dust will be contained in either normal star-forming regions or large dusty star-forming clumps, and the ISM will have little dust. Therefore, since both types of star-forming regions are expected to be roughly spherical, inclination dependence in either the nebular or stellar reddening will be minimal. To test whether this also holds for the galaxies in this study, we plot the axis ratio (a proxy for inclination) against the nebular reddening from the Paschen lines and the stellar reddening from the SED fitting in Figure~\ref{f:b_a_red}. We see no significant correlation between the axis ratio and the nebular or stellar reddening, implying that both the stellar and nebular light are being attenuated by spherically symmetric dust distributions. In conjunction with the trend that more massive galaxies have slightly elevated differential reddening than low-mass galaxies, our results support the three-component dust model in these galaxies.

\section{Metallicity and Attenuation curves}\label{s:metallicity}

In this section, we focus on how the choice of nebular attenuation curve affects gas-phase metallicity measurements using the strong-line method. This method involves measurements of strong line ratios and comparisons to calibrations measured in other works. Here we focus on the calibrations presented in \citet{sandersAURORASurveyHighRedshift2025} using spectroscopy from the AURORA survey. Because these diagnostics cover a wide redshift range that includes our sample, these calibrations are more appropriate than the often-used local relations presented in \citet{curti_mass-metallicity_2020}. 

This section does not aim to measure the true gas-phase metallicity for each galaxy, but to highlight how important variability in the nebular attenuation curve is for different diagnostics. Because some diagnostics use emission lines with very small wavelength differences and others use more widely spaced emission lines, the choice of attenuation curve will affect each diagnostic differently.

We present the diagnostics used in this work in Table~\ref{t:metal_diags}, based on \citet{sandersAURORASurveyHighRedshift2025}. In this work, we use any diagnostic that includes only emission lines within the wavelength range of the nebular attenuation curves we measure ($4102<\lambda_{\rm rest}<12819$\,\AA). We also include the normalised wavelength range ($\Delta\lambda/\left<\lambda\right>$) in Table~\ref{t:metal_diags}, defined as the range of emission-line wavelengths used in each diagnostic divided by the average wavelength of the emission lines in that diagnostic.

\begin{table}
  \centering
    \begin{tabular}{lll}\hline
        Notation & Line Ratio & $\Delta\lambda/\left<\lambda\right>$  \\\hline
        O3 & $\rm [O~III]\lambda5008/H\beta$ & 0.030 \\
        N2 & $\rm [N~II]\lambda 6585/H\alpha$ & 0.003\\
        O3N2 & $\rm ([O~III]\lambda 5008/H\beta)/([N~II]\lambda 6585/H\alpha)$ & 0.300 \\
        N2S2 & $\rm [N~II]\lambda 6585/[S~II]\lambda\lambda 6718,6733$ & 0.022 \\
        S3 & $\rm [S~III]\lambda\lambda 9071,9533/H\alpha$ & 0.354 \\
        S2 & $\rm [S~II]\lambda\lambda 6718,6733/H\alpha$ & 0.025\\
        S23 & $\rm ([S~III]\lambda\lambda 9071,9533 + [S~II]\lambda\lambda 6718,6733)/H\alpha$ & 0.385\\
        S32 & $\rm [S~III]\lambda\lambda 9071,9533 / [S~II]\lambda\lambda 6718,6733$ & 0.351 \\
        S3O3 & $\rm [S~III]\lambda\lambda 9071,9533 / [O~III]\lambda5008$ & 0.575 \\
        O3S2 & $\rm ([O~III]\lambda5008/H\beta)/([S~II]\lambda\lambda 6718,6733/H\alpha)$ & 0.313 \\
        Ar3 & $\rm [Ar~III]\lambda 7137/H\alpha$ & 0.084 \\
        Ar3O3 & $\rm [Ar~III]\lambda 7137/[O~III]\lambda 5008$ & 0.351 \\ \hline
    \end{tabular}\\[5pt]
    \medskip
    \caption{Table showing definitions of the emission-line ratios used to determine galaxy metallicity through strong-line calibrations, taken from \citet{sandersAURORASurveyHighRedshift2025}. The normalised wavelength range of the emission lines in each diagnostic ($\Delta\lambda/\left<\lambda\right>$) is the range of emission-line wavelengths divided by the average wavelength in each diagnostic. The notation $\rm [S~II]\lambda\lambda 6718,6733$ is equivalent to writing $\rm [S~II]\lambda 6718 + [S~II]\lambda 6733$.}
    \label{t:metal_diags}
\end{table}

To measure the metallicity of each galaxy, we identify which diagnostics apply by setting a signal-to-noise limit of $3~\sigma$ on every emission line used in the diagnostic. Then, to dust-correct the emission lines used in each diagnostic, we either use our measured nebular attenuation curve or the \citet{calzetti_dust_2000} attenuation curve. We describe how this is performed below. 

The nebular attenuation curve we measure in this work is relative (as discussed in depth in Section~\ref{s:calc_atten}), so we cannot dust-correct a single emission line, which requires the absolute attenuation at a specific wavelength. However, we can dust-correct emission line ratios, since this requires only the relative attenuation between two wavelengths. Therefore, we separate out each diagnostic into a series of individual emission line ratios and dust-correct each line ratio using the following equation:

\begin{align}\label{eq:line_rat_corr}
    \frac{F_a^i}{F_b^i} = \frac{F_a}{F_b}10^{0.4[A'(\lambda_a)-A'(\lambda_b)]}
\end{align}

where $F_a^i/F_b^i$ is the intrinsic, dust-corrected line ratio, $F_a/F_b$ is the observed line ratio, and $A'(\lambda_a)-A'(\lambda_b)$ is the difference in attenuation between the wavelengths of the two emission lines. 

When we instead used the \citet{calzetti_dust_2000} attenuation curve to dust-correct the emission line ratios, we used the Balmer decrement (H$\alpha$/H$\beta$) to obtain $A(\lambda_a)-A(\lambda_b)$ (which is equal to $A'(\lambda_a)-A'(\lambda_b)$) to input into Equation~\ref{eq:line_rat_corr}. 

Since we fit all emission-line fluxes and nebular attenuation curves in this work using MCMC methods, we have full posteriors for these measurements. Therefore, to measure the uncertainty on the intrinsic line ratios, and thus on the metallicity from each diagnostic, we randomly sample the posterior of each emission line and the attenuation curve 1000 times to build up a posterior on the final metallicity. This applies to both methods of dust correcting the emission lines.

Because many strong-line diagnostics presented in \citet{sandersAURORASurveyHighRedshift2025} are double-valued, and the metallicity range over which each diagnostic is valid varies, we show below how we select each solution. 

First, we discard any solution that yields a metallicity outside the range $7.0<12+\log(\rm O/H)<8.8$. Although this range is wider than the valid calibration range presented in \citet{sandersAURORASurveyHighRedshift2025}, each calibration behaved reasonably within it, and for the purpose of exploring how the attenuation curve affects metallicity measurements, extrapolation into this range is valid. Then, for any double-valued metallicity within this range, we keep both solutions, since we are not interested in the absolute value of the metallicity, only how the shape of the attenuation curve affects the measurement. We note if a double-valued solution is included by adding "$\_1$" next to the diagnostic where relevant. 

\begin{figure*}
\begin{tabular}{cc}
    \includegraphics[width=0.7\textwidth]{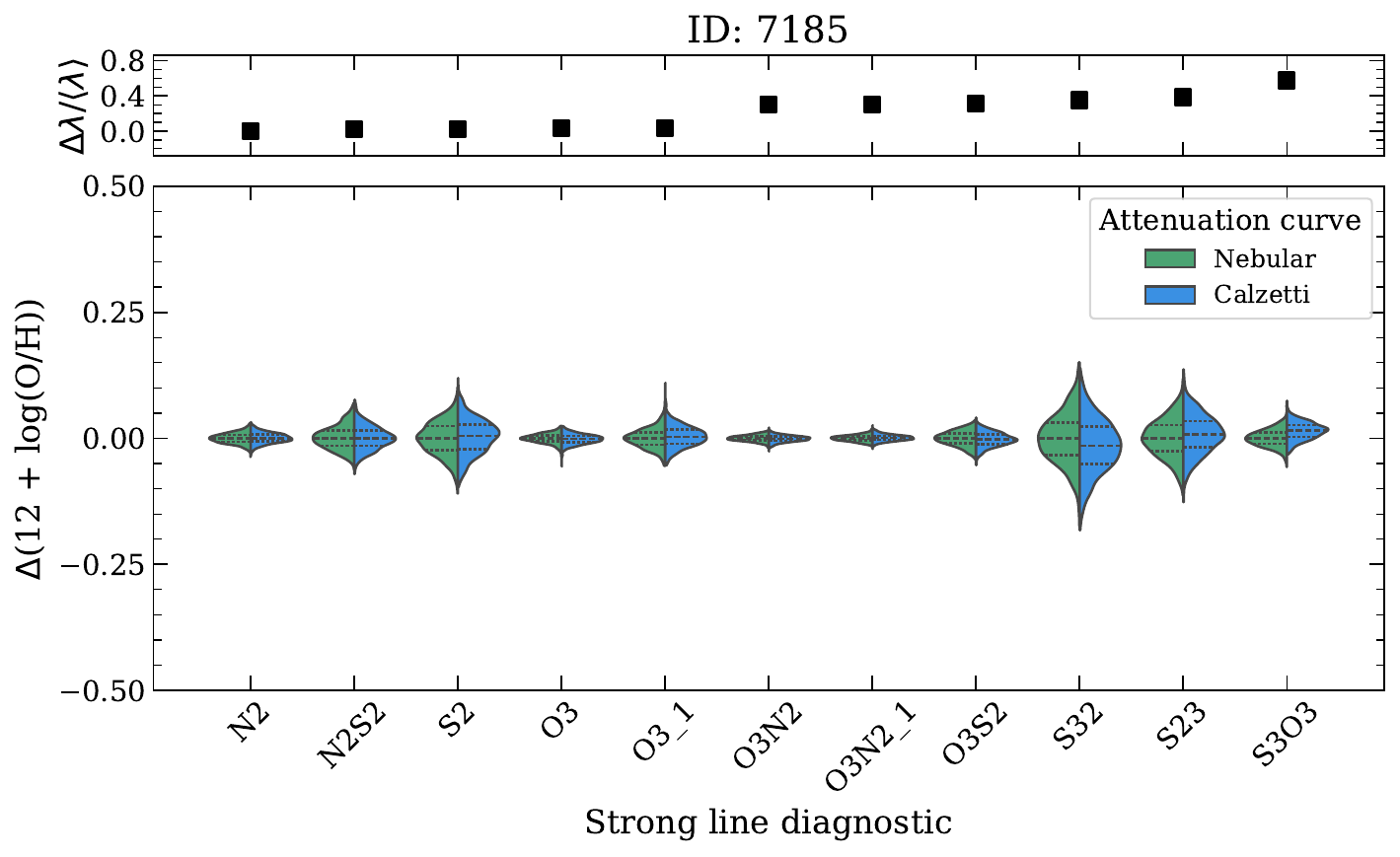} \\
    \includegraphics[width=0.7\textwidth]{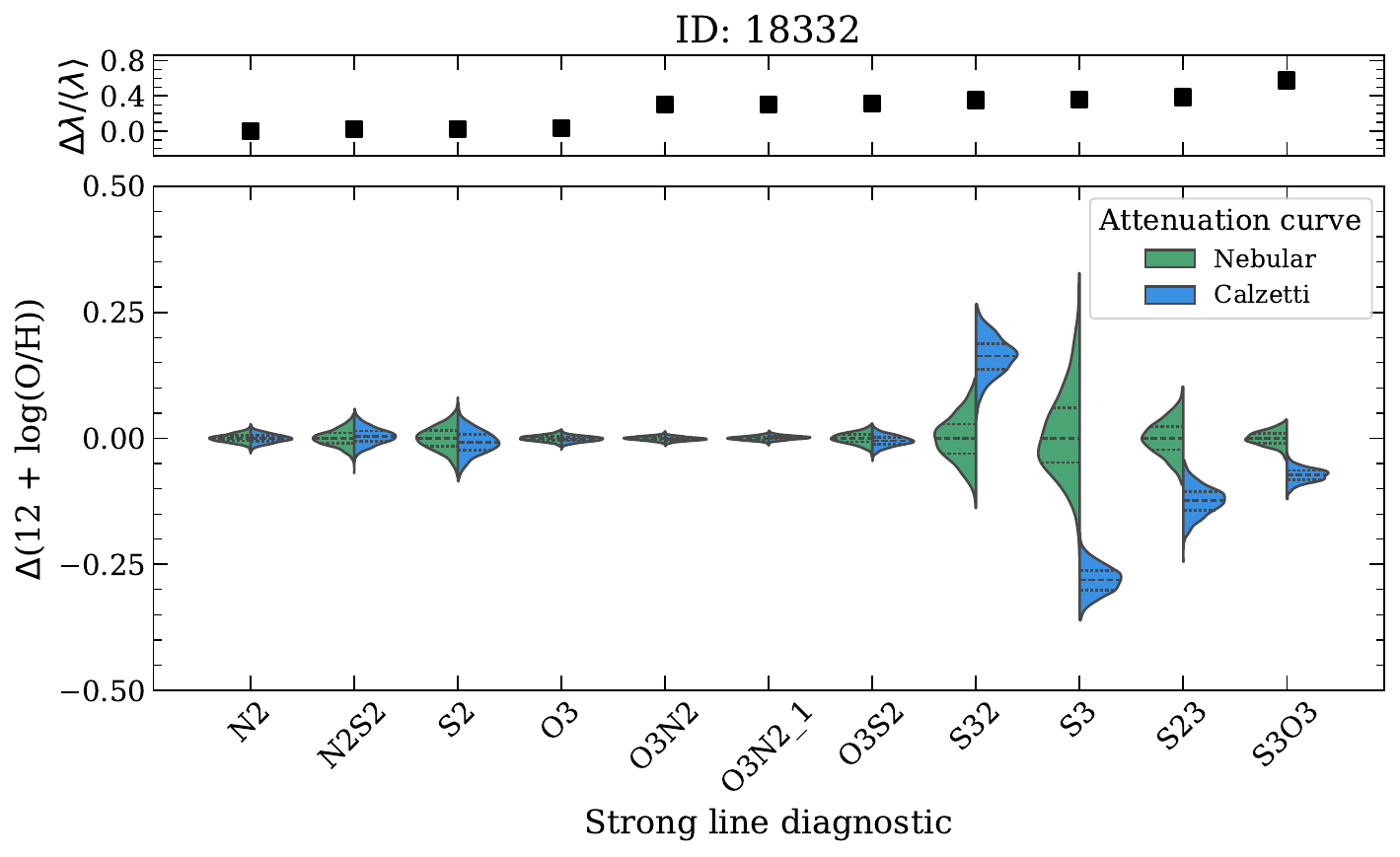} \\ 
\end{tabular}
\caption{Violin plot showing the posterior distribution of metallicity using the nebular attenuation curve measured in this work (green) or the \citet{calzetti_dust_2000} attenuation curve (blue) for each available diagnostic, for two example galaxies. For better comparison, we have shifted every distribution such that the distribution using the measured nebular attenuation curve is centred on zero. In the top panel of each plot, we show the normalised wavelength range of the emission lines used in each diagnostic ($\Delta\lambda/\left<\lambda\right>$), quantifying how sensitive the diagnostic should be to dust attenuation. For the galaxy with ID 7185, the metallicity measurements show no significant difference between the two attenuation-curve choices for any diagnostic. This is because the attenuation curve closely matches the \citet{calzetti_dust_2000} curve, and the Balmer decrement well represents the galaxy's total reddening. However, for the galaxy with ID 18332, several diagnostics report significantly different metallicities depending on the choice of attenuation curve, namely the O3, S32, S23 and S3O3 diagnostics. This is because the best-fit attenuation curve is shallower than the \citet{calzetti_dust_2000} curve, and the Balmer decrement overestimates the total reddening for this galaxy as described in the text. The difference in metallicity can be as large as $\sim 0.3$ dex if the wrong attenuation curve is chosen.} \label{f:metal_violin}
\end{figure*}

In Figure~\ref{f:metal_violin}, we compare metallicity measurements for each diagnostic using the nebular attenuation curve measured in this work versus the \citet{calzetti_dust_2000} curve for two example galaxies. These two galaxies were selected as examples where the different attenuation curves lead to either little or significant difference in the measured metallicity. For each diagnostic, we present a violin plot of the posterior distribution of the metallicity measurement using each attenuation curve. We order the diagnostics by their normalised wavelength ranges ($\Delta\lambda/\left<\lambda\right>$), which measures how spread the emission lines are and whether they are centred at short or long wavelengths. This quantifies how affected each diagnostic should be by dust attenuation, and thus by the choice of attenuation curve. We show this normalised wavelength range in the top panel of each figure.

For galaxy 7185, the metallicity distribution shape varies between diagnostics (due to the varying S/N of the emission lines used in each diagnostic), but the attenuation curve used does not significantly affect the distribution averages. This is because the measured nebular attenuation curves for these galaxies are very similar in shape to the \citet{calzetti_dust_2000} curve, and the Balmer decrements well represent their total dust attenuation.

However, in galaxy 18332, some diagnostics yield significantly different metallicity measurements, with discrepancies as high as $\sim 0.3$ dex. This is because the nebular attenuation curve we measure in this work differs significantly from that of \citet{calzetti_dust_2000} (see Figure~\ref{f:k_lam_grid}), or because the Balmer decrement poorly approximates the galaxy's overall reddening.

This difference appears only in some diagnostics for these galaxies, while several diagnostics agree regardless of the attenuation curve used. The diagnostics most affected by the choice of attenuation curve typically have a large normalised wavelength separation; thus, the attenuation curve shape becomes important. Less-affected diagnostics typically use emission lines close in wavelength or centred at longer wavelengths, where attenuation is weaker. However, O3S2 is an exception. Although it has a relatively large normalised wavelength range because it combines several widely spaced emission lines, each individual line ratio in this diagnostic has a small wavelength range and is thus less affected by dust attenuation. 

\begin{figure*}
\centerline{\includegraphics[width=.8\textwidth]{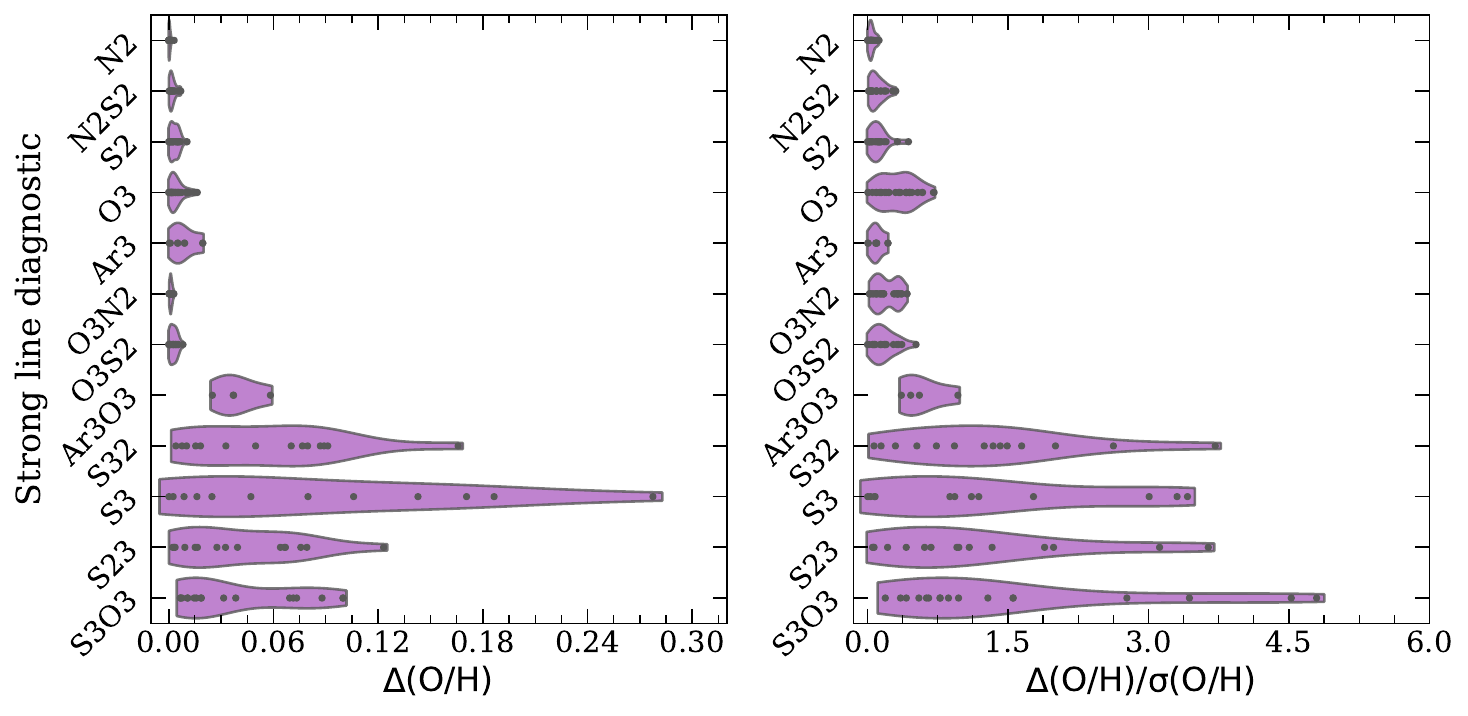}}
\medskip
\caption{Histogram of the difference in the measurement of metallicity when using the nebular attenuation curve derived in this work and the \citet{calzetti_dust_2000} attenuation curve ($\Delta$(O/H), left), and this difference divided by the error on the measurement using the nebular attenuation curve ($\Delta$(O/H)/$\sigma$(O/H), right), for each diagnostic used for any galaxy in the sample. The distribution of $\Delta$(O/H) shows the magnitude of the metallicity difference, and the distribution of $\Delta$(O/H)/$\sigma$(O/H) probes how this difference compares to the metallicity measurement uncertainty. Measurements using the S32, S3, S23, and S3O3 diagnostics can differ by over $0.1$ dex, or $1~\sigma$, and as much as $0.3$ dex, or $5~\sigma$; therefore, the uncertainty from the choice of attenuation curve is significantly larger than the measurement uncertainty. Other diagnostics, namely N2, N2S2, S2, O3, Ar3, O3N2, O3S2, and Ar3O3, are always less than $0.1$ dex, or $1~\sigma$, so the uncertainty introduced by the choice of attenuation curve is no larger than the measurement uncertainty.} \label{f:metal_sig}
\end{figure*}

To quantify which diagnostics are most affected by the choice of attenuation curve across all the galaxies, we calculate the median of the metallicity distribution for each galaxy and each diagnostic using the nebular and the \citet{calzetti_dust_2000} attenuation curve, then take the difference in median to obtain $\Delta$(O/H). This quantifies for which diagnostics the measured metallicity difference can be large or small, and we show its distribution in the left panel of Figure~\ref{f:metal_sig}. Dividing this difference by the standard deviation of the metallicity from the nebular curve gives $\Delta \rm (O/H) / {\sigma}(O/H)$, which probes whether the uncertainty introduced by the choice of attenuation curve exceeds the measurement uncertainty on the metallicity itself. We show the distribution of this in the right panel of Figure~\ref{f:metal_sig}.

For some diagnostics, such as S32, S3, S23, and S3O3, the differences range from zero to $0.3$ dex, or over $5~\sigma$, implying that an incorrect choice of attenuation curve can introduce errors significantly larger than the measurement uncertainty. These diagnostics should therefore be used with caution if the nebular attenuation curve cannot be measured. However, for diagnostics such as N2, N2S2, S2, O3, Ar3, O3N2, O3S2 and Ar3O3, the differences are always less than $0.1$ dex, or within $1~\sigma$, and so the uncertainty introduced by the incorrect choice of the attenuation curve is not significantly larger than the measurement uncertainty on the metallicity. Thus, for these latter diagnostics, using the Balmer decrement and assuming the \citet{calzetti_dust_2000} attenuation curve will give reliable metallicities. 

Here, we consider the impact on metallicity using only the \citet{calzetti_dust_2000} curve or the nebular curve derived in this work. As Section~\ref{s:comp_studies} shows, for most galaxies the nebular attenuation curve aligns closely with the Milky Way extinction curve and is slightly shallower than the \citet{calzetti_dust_2000} curve, yet these small differences, along with how well the Balmer decrement represents each galaxy's overall dust attenuation, can still lead to significant differences in the measured metallicities. Therefore, using a curve that diverges more from the \citet{calzetti_dust_2000} curve, such as the average \citet{reddyAURORASurveyMultiple2026a} curve, would yield even larger metallicity differences for the galaxies in this sample.

This analysis shows that the uncertainty introduced by choosing an incorrect attenuation curve to correct the emission lines can now exceed the measurement uncertainty on these lines. It is therefore crucial that we obtain larger studies that measure the nebular attenuation curve to trace the main driver of variation between galaxies, so we can accurately constrain properties such as gas-phase metallicity. Until then, we urge caution when using strong-line calibrations to measure metallicity and recommend using only diagnostics with narrow wavelength spreads in the emission lines, since even a high signal-to-noise measurement of the Balmer decrement can lead to incorrect metallicities if an incorrect nebular attenuation curve shape is assumed.

\section{Conclusions}\label{s:conclusion}

In this work, we use high signal-to-noise spectra from JWST/NIRSpec for 24 galaxies ($9.3<\log(\rm M{\star}~[M_\odot])<11.1$) at cosmic noon ($1.8<z<3.3$), selected from the Blue Jay survey. These galaxies have at least five $5\sigma$ detections of rest-optical to rest-NIR hydrogen recombination lines, from H$\delta$ to Pa$\beta$, allowing us to determine individual nebular attenuation curves. 

We take the ratio of each hydrogen recombination line to H$\alpha$ and compare it to the intrinsic line ratio under Case B recombination to measure the nebular attenuation curves, which show little diversity. We then leverage the SED fits of these galaxies, using JWST/NIRCam photometry and NIRSpec spectra with the \texttt{Prospector} fitting code, to constrain stellar population properties and measure the stellar continuum attenuation curve and stellar reddening. We compare this stellar reddening with our measurement of the nebular reddening, which is unaffected by the systematic uncertainty associated with the choice of nebular attenuation curve. Finally, we explore how the choice of attenuation curve affects the measured metallicity. The main results of this study are as follows:

\begin{enumerate}
    \item We observe minimal diversity in the shapes of the nebular attenuation curves, with many of the curves close in slope to that of the Milky Way extinction curve \citep{cardelli_relationship_1989} and slightly shallower than the \citet{calzetti_dust_2000} attenuation curve.
    \item Comparing nebular reddening from the Balmer lines with that from the Paschen lines shows that for $90\%$ of our sample, the two measures agree within $1\sigma$, indicating that obscured star formation does not significantly affect these galaxies.
    \item We find that our median nebular attenuation curve is significantly shallower than the average curve from \citet{reddyAURORASurveyMultiple2026a}, a similar study at similar redshifts, and that their curves are significantly more diverse than ours. This dichotomy could reflect differences in galaxy properties driven by different sample selection techniques. 
    \item We then extracted the stellar reddening from the \texttt{Prospector} SED fits and compared this to the nebular reddening we measured from the nebular attenuation curves. We find that the nebular reddening is, on average, 1.91 times larger than the stellar reddening, $1~\sigma$ below the canonical value of 2.27 from \citet{calzetti_dust_2000}, and consistent with the nebular reddening being, on average, greater than the stellar reddening. 
    \item We then measure the differential reddening ($\rm \Delta E(B-V) \equiv E(B-V)_{neb} - E(B-V)_{stellar}$) to explore how the difference in nebular and stellar reddening relates to the galaxy properties. We find that it increases with stellar mass, but shows no trend with SFR. This increase with stellar mass is consistent with recent work covering this redshift range \citep{karthikeyanBalmerDecrementsNebularStellar2026}. Combined with our result that the stellar and nebular reddening do not correlate with the axis ratio (a proxy for inclination), our results support the three-component dust model proposed by \citet{lorenz_updated_2023}.
    \item We then use the derived nebular attenuation curves and the commonly assumed \citet{calzetti_dust_2000} attenuation curve to assess the biases introduced when measuring gas-phase metallicity through the strong line method. 
    For diagnostics that use emission lines with a large wavelength spread, or lines centred in the rest-UV, the choice of attenuation curve can introduce biases up to 5 $\sigma$, particularly for the S32, S3, S23, and S3O3 diagnostics. We therefore recommend avoiding these diagnostics where possible, since even a high signal-to-noise measurement of the Balmer decrement can lead to incorrect metallicities if the incorrect nebular attenuation curve shape is assumed. 
\end{enumerate}

These results are made possible by high signal-to-noise JWST/NIRSpec spectra from the Blue Jay survey \citep{belliBlueJaySurvey2025} and the methodology to measure nebular attenuation curves developed by \citet{reddy_mosdef_2020}, \citet{sanders_aurora_2024}, and \citet{reddyAURORASurveyMultiple2026a}. In this work, we find negligible diversity in the nebular attenuation curves, whereas \citet{reddyAURORASurveyMultiple2026a} find significant diversity. To understand what causes this difference, and to identify the main driver of diversity in attenuation-curve shapes, we need larger studies that measure nebular attenuation curves for individual galaxies across redshift. By better understanding how nebular attenuation curves vary between galaxies, we can remove a large, currently unconstrained source of uncertainty when measuring key galaxy measurements.

\section*{Data Availability}

The JWST/NIRSpec spectra and JWST/NIRCam imaging used in this paper were obtained through the Cycle 1 program Blue Jay (GO 1810, PI Belli) and are publicly available for download from the MAST archive at https://dx.doi.org/10.17909/sgjm-gq64.



\bibliographystyle{mnras}
\bibliography{references_new} 





\appendix

\section{Measuring Electron Temperature and Density}\label{app:electron_t_n}

Here we show how we calculate the electron temperature ($T_e$) and electron density ($n_e$) in the HII regions of our galaxies using various measured emission lines using the Python package \texttt{Pyneb} \citep{luridianaPyNebNewTool2015}, and how this has a negligible impact on the value of the intrinsic hydrogen line ratios. 

The conditions of the region from which emission lines originate, such as the electron density and temperature, are imprinted on the emission lines \citep{draine_physics_2011}. Some emission lines are sensitive to electron density but vary little with electron temperature, such as the [SII]$~\lambda 6717,31$ and [OII]$~\lambda 3726,29$ doublets. A set of emission lines, referred to as the "auroral" emission lines, is sensitive to the electron temperature but varies little with the electron density. Combining measurements of the density and temperature-sensitive emission lines allows us to constrain the electron temperature and density. For a detailed overview of how these different emission lines relate to the electron density and temperature, we refer the reader to \cite{maiolinoReMetallicaCosmic2019}.

In our medium-resolution NIRSpec spectra of the Blue Jay survey, we can resolve the [SII]$~\lambda 6717,31$ doublet; however, we cannot resolve the [OII]$~\lambda 3726,29$ doublet, so we do not consider it in this work. Various auroral lines were detected in each galaxy considered in this work, such as the [SIII]$~\lambda 6312, 9069$, [OIII]$~\lambda 4363$ and [NII]$~\lambda 5755$ emission lines. These emission lines are also used to measure the gas-phase metallicity for the full Blue Jay sample, which is presented in Bugiani et al. (in prep.).

\begin{figure}
\centerline{\includegraphics[width=.5\textwidth]{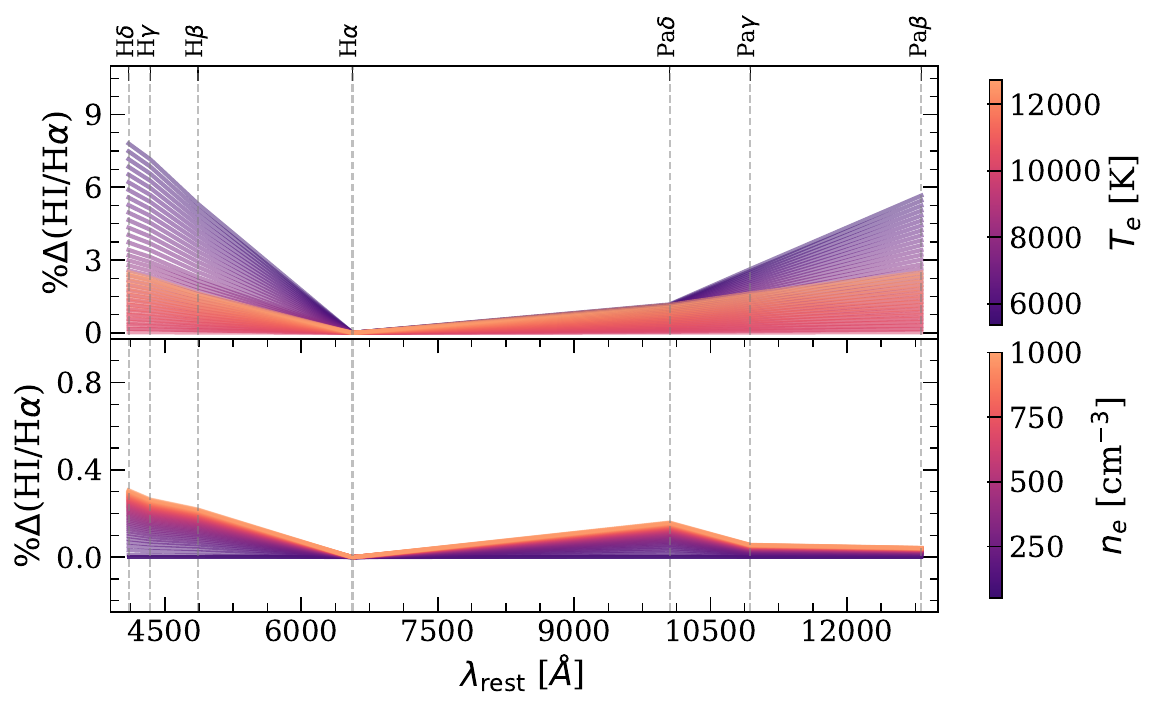}}
\medskip
\caption{Percentage deviation of the intrinsic line ratios from the ratio at $T_e=10,000$~K and $n_e=100$~cm$^{-3}$ of seven hydrogen recombination lines (indicated at the top) with respect to H$\alpha$ as the electron temperature ($T_e$) in the top panel and the electron density ($n_e$) in the bottom panel are varied. As $T_e$ is varied, $n_e$ is fixed to $100$~cm$^{-3}$, and as $n_e$ is varied, $T_e$ is fixed to $10,000$~K.} \label{f:Te_ne_var}
\end{figure}

Using \texttt{Pyneb}, we calculate the electron temperature and density for the galaxies, averaging the electron densities and temperatures if more than one auroral line is detected. These measurements yield a range of electron temperatures and densities for our sample of galaxies, with electron temperatures varying between $5,000-12,000$~K and electron densities between $200-1100 \rm ~ cm^{-3}$. Although these ranges are broad, they have little impact on the intrinsic hydrogen recombination line ratios. To demonstrate this, we calculate the line ratio using a fiducial electron temperature ($T_e = 10,000$~K) and density ($n_e = 100$~cm$^{-3}$), typical nebular conditions assumed in many works, e.g., \citet{reddyPaschenlineConstraintsDust2023}.  We then calculate the line ratio by varying the electron temperature across the range $5,000-12,000$~K while keeping the electron density fixed at the fiducial value, and compare it with the fiducial line ratio. We then repeat this for the electron density, varying it across the range $50-1,000$~cm$^{-3}$, and plot the results in Figure~\ref{f:Te_ne_var} for each line ratio. The largest variation among the line ratios is H$\delta$/H$\alpha$ with electron temperature, with a maximum variation of about $8\%$, which is not significant compared to the uncertainty on the emission line ratios. The variation with electron density is negligible.

For some galaxies, we could not determine the electron density and temperature because the necessary emission-line S/N was insufficient. For consistency, and since the intrinsic line ratios do not vary significantly with electron density and temperature, we use the standard values of the electron temperature ($T_e = 10,000$~K) and density ($n_e = 100$~cm$^{-3}$).

The electron temperatures and densities are then used to calculate the intrinsic line ratio between the hydrogen recombination lines used in this work and H$_{\alpha}$. The intrinsic line ratios with H$\alpha$ given the average $T_e$ and $n_e$ measured here are shown in Table~\ref{t:int_ratios}.

\section{Median attenuation curve comparisons}\label{app:med_atten_comp}

\begin{figure}
\centerline{\includegraphics[width=.5\textwidth]{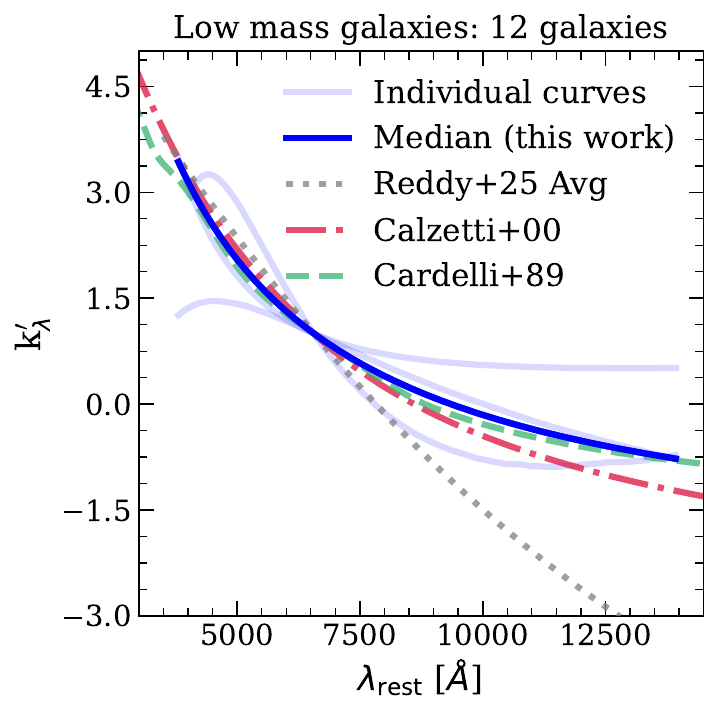}}
\medskip
\caption{The median nebular attenuation curve($k'(\lambda)$, in dark blue), plotted next to the individual galaxy attenuation curves (in pale blue), however, now only for the low stellar mass galaxies with $\log(\rm M_\star/M_\odot)<10.07$ (12 galaxies, 50\% of the sample). We also plot the average nebular attenuation curve from \citet{reddyAURORASurveyMultiple2026a} as the grey dotted line, the \citet{calzetti_dust_2000} attenuation curve for local starbursts as the red dashed-dot line, and the \citet{cardelli_relationship_1989} Milky Way extinction curve as the green dashed line, all normalised such that the attenuation at $H\alpha$ is unity. There is little diversity, and the median aligns well with the Milky Way extinction curve.} \label{f:k_lam_median_low_mass}
\end{figure}

\begin{figure}
\centerline{\includegraphics[width=.5\textwidth]{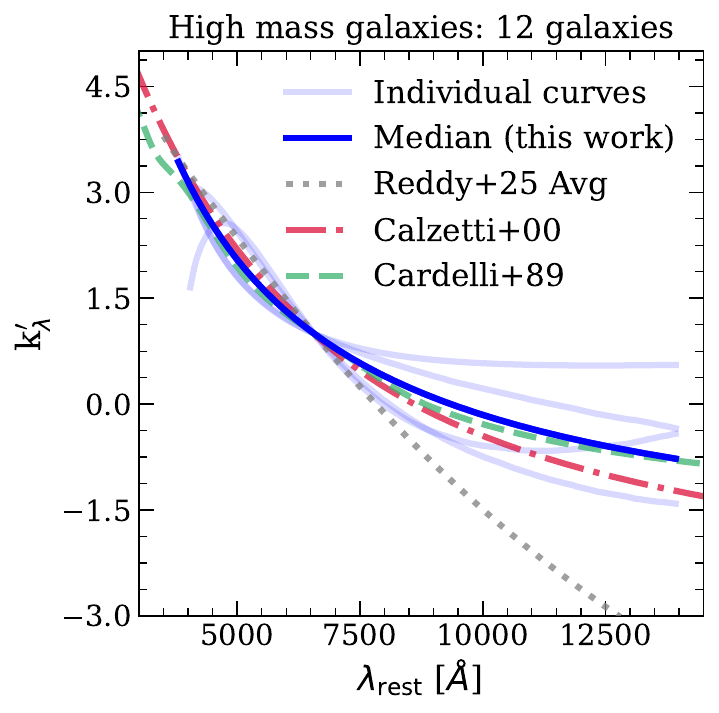}}
\medskip
\caption{The same as Figure~\ref{f:k_lam_median_low_mass} but for only the high stellar mass galaxies with $\log(\rm M_\star/M_\odot)>10.07$ (12 galaxies, 50\% of the sample). There is little diversity, and the median aligns well with the Milky Way extinction curve, although there is more diversity than for the low-mass sample.} \label{f:k_lam_median_high_mass}
\end{figure}

\begin{figure}
\centerline{\includegraphics[width=.5\textwidth]{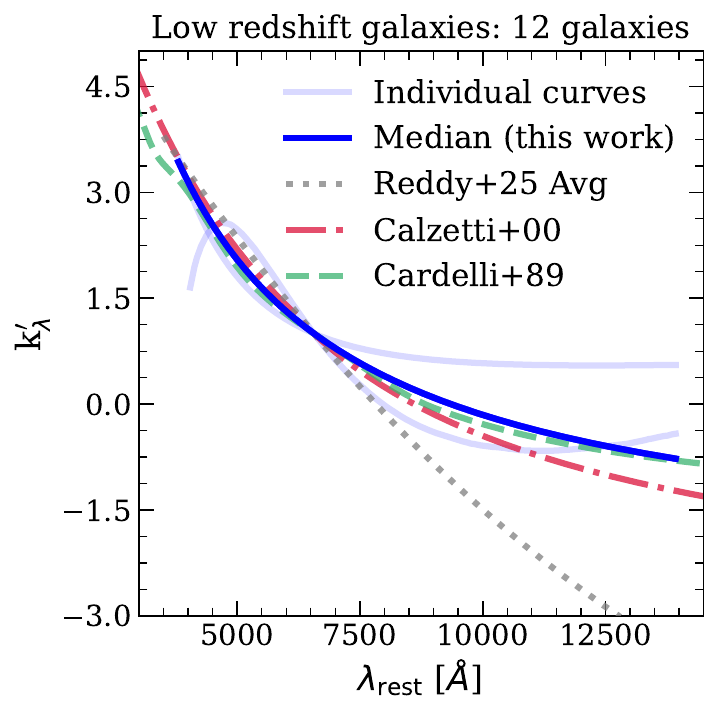}}
\medskip
\caption{The same as Figure~\ref{f:k_lam_median_low_mass} but for only galaxies with $z<2.5$ (12 galaxies, 50\% of the sample). There is little diversity, and the median aligns well with the Milky Way extinction curve.} \label{f:k_lam_median_low_z}
\end{figure}

\begin{figure}
\centerline{\includegraphics[width=.5\textwidth]{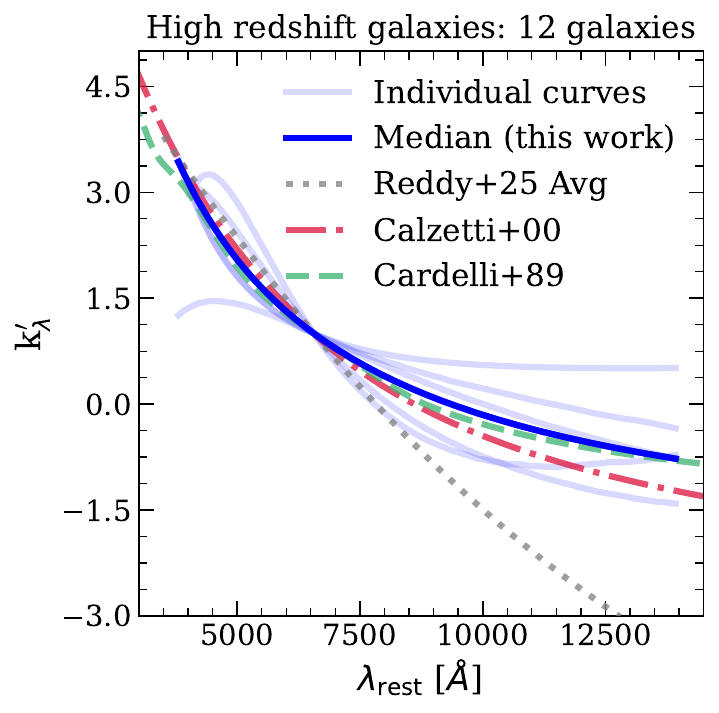}}
\medskip
\caption{The same as Figure~\ref{f:k_lam_median_low_mass} but for only galaxies with $z>2.5$ (12 galaxies, 50\% of the sample). There is some diversity, and although the median aligns well with the Milky Way extinction curve, there is more diversity than for the low-redshift sample.} \label{f:k_lam_median_high_z}
\end{figure}

Here we show how the median attenuation curve derived in this work for the sample split first in stellar mass (Figure~\ref{f:k_lam_median_low_mass}-\ref{f:k_lam_median_high_mass}) then in redshift (Figure~\ref{f:k_lam_median_low_z}-\ref{f:k_lam_median_high_z}), ensuring an even split in both. 

\section{Nebular reddening extracted from SED fitting}\label{app:neb_sed}

\begin{figure}
\centerline{\includegraphics[width=.5\textwidth]{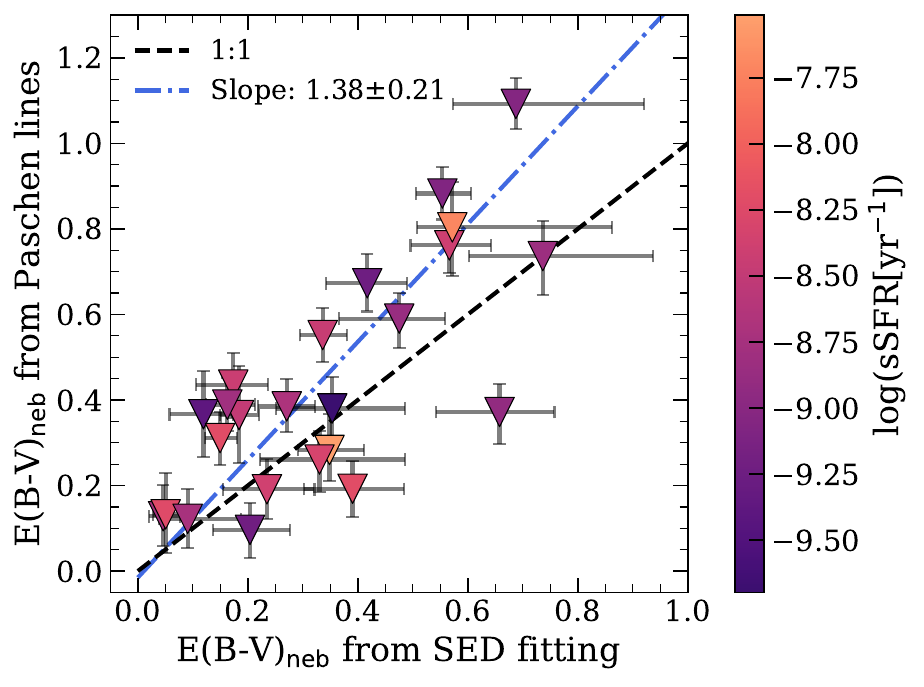}}
\medskip
\caption{The nebular reddening, E(B-V)$_{\rm neb}$, measured from the Paschen lines, plotted against the nebular reddening measured from SED fitting, colour-coded by the sSFR. These two measures of the nebular reddening are largely consistent with each other, with many galaxies lying close to the 1:1 line (black dashed line). This shows that the SED fitting is modelling the nebular reddening well.} \label{f:red_neb_neb_SED}
\end{figure}

Here we compare the nebular reddening measured in this work (see Section~\ref{s:obsc_sf}) to the nebular reddening extracted from the \texttt{Prospector} SED fits. Because we use a two-component dust model in the SED fit, we can calculate the nebular attenuation curve by combining the ISM dust and birth-cloud dust attenuation curves. This, of course, comes with many caveats, since we assume the dust model, dust attenuation curve shapes, and several other important properties of the stellar population (see Section~\ref{s:sed_fit}); however, it is still interesting to compare these results. In Figure~\ref{f:red_neb_neb_SED} we plot the nebular reddening derived from the Paschen lines against that derived from the SED fitting. They generally agree, showing that for these galaxies, the SED fitting constrains the dust model well.

\bsp	
\label{lastpage}
\end{document}